\documentclass[prd,a4paper,
twocolumn,amsmath, amssymb, floatfix,
nofootinbib,wrapfloat byrevtex]{revtex4-2}

 \usepackage{amsmath,mathrsfs,upgreek,graphicx,placeins,float}
 \usepackage{color}
 \usepackage{booktabs}
 \usepackage{multirow}
 
 \numberwithin{equation}{section}

\usepackage[latin9]{inputenc}

\usepackage{amsmath}
\usepackage{amsfonts}
\usepackage{bbm}
\usepackage{microtype}
\usepackage{hyperref}
\usepackage{cleveref}
\usepackage{color}
\usepackage{mathrsfs}
\usepackage{amsthm,amsmath,amssymb}
\usepackage{upgreek}

\makeatletter
\DeclareRobustCommand{\rcite}[1]{%
  \rcite@aux#1,\@nil{#1}%
}
\def\rcite@aux#1,#2\@nil#3{%
  \if\relax#2\relax
    Ref.~\cite{#3}%
  \else
    Refs.~\cite{#3}%
  \fi
}
\makeatother

\begin{document}
\title{Reformulation and Extension of the Standard Model}

\author{Peng Huang}
\email{hp@zcmu.edu.cn}
\affiliation{Physics Teaching and Research Office, Zhejiang Chinese Medical University, \\
 Hangzhou 310053, China}

\begin{abstract}
We present a classically equivalent reformulation of the Standard Model. In this framework, the Higgs doublet is recast as a $2\times2$ matrix and right-handed fermion singlets are organized into novel doublets. This restructuring reveals a latent algebraic geometry that naturally realizes a new local gauge principle: the \textbf{extended Weyl symmetry}. Generalizing Hermann Weyl's 1929 idea of local scale invariance to internal multiplet spaces, this symmetry extends the scope of local gauge symmetries beyond the conventional Yang--Mills framework and provides the foundation for a classical extension of the Standard Model. The resulting theory introduces new gauge bosons (potential dark matter candidates) and a second scalar field. Weyl symmetry breaking renders this additional scalar non-dynamical and leaves the observed Higgs boson as the sole dynamical scalar. The extended Weyl symmetry also leads to several notable theoretical consequences: (i) parity violation is embedded within an underlying parity-symmetric structure; (ii) the minus sign of the Higgs potential is uniquely fixed by the symmetry breaking pattern rather than chosen \textit{by hand}; and (iii) a right-handed neutrino is required by the doublet structure, with its masslessness protected by the extended Weyl symmetry. Quantum aspects of this framework---particularly potential gauge anomalies and renormalizability---remain open and warrant dedicated investigation.\end{abstract}
\preprint{}

\maketitle

\section{Introduction}\label{introduction}

Observe that the Standard Model (SM) \cite{Glashow:1961tr,Weinberg:1967tq,Salam:1968rm} admits an equivalent---yet more compact---reformulation. Its most fundamental mathematical content is captured by the following identity:
\begin{equation}
\label{807}
\begin{aligned}
&\ \ \ \ \textbf{Q}_L^{\dagger}\textbf{H}d_R+\textbf{Q}_L^{\dagger}\widetilde{\textbf{H}}u_R\\
&=\begin{pmatrix}u_L^{\dagger}&d_L^{\dagger}\end{pmatrix}\begin{pmatrix} \Phi_1\\ \Phi_2 \end{pmatrix}d_R+\begin{pmatrix}u_L^{\dagger}&d_L^{\dagger}\end{pmatrix}\begin{pmatrix} \Phi_2^*\\ -\Phi_1^* \end{pmatrix}u_R\\
&=\textbf{Q}_L^{\dagger}(\textbf{H}'+\widetilde{\textbf{H}}')\textbf{Q}_R\\
&=\begin{pmatrix}u_L^{\dagger}&d_L^{\dagger}\end{pmatrix}\Big{(}\begin{pmatrix}0& \Phi_1\\ 0& \Phi_2 \end{pmatrix}+\begin{pmatrix}\Phi^*_2& 0\\ -\Phi^*_1& 0 \end{pmatrix}\Big{)}\begin{pmatrix}u_R\\d_R\end{pmatrix}.
\end{aligned}
\end{equation}

Here, $\textbf{Q}_L = \begin{pmatrix} u_L & d_L \end{pmatrix}^{\!T}$ is the $SU(2)_L$ doublet formed by the left-handed up and down quarks ($u_L$ and $d_L$); $u_R$ and $d_R$ are the right-handed up and down quarks, which are $SU(2)_L$ singlets; $\textbf{H} = \begin{pmatrix} \Phi_1 & \Phi_2 \end{pmatrix}^{\!T}$ is the Higgs doublet, and $\widetilde{\textbf{H}}=\epsilon\textbf{H}^*$ is its charge-conjugated field. It is crucial to emphasize that $\textbf{Q}_R = \begin{pmatrix} u_R & d_R \end{pmatrix}^{\!T}$ is a \textit{newly assembled doublet} requiring careful definition, whose specific meaning will be elaborated in Sec.~\ref{jia}; the compactness of this formulation stems precisely from assembling the originally $SU(2)_L$ inert singlets into such doublets. Likewise, the newly introduced matrices $\textbf{H}'$ and $\widetilde{\textbf{H}}'$ require careful definition and will be elaborated in Sec.~\ref{jia}.

At first glance, Eq.(\ref{807}) appears merely a rearrangement of terms. In the standard Yukawa form (second line), all entries are fixed by known fields, leaving no apparent room for additional DoFs.
Yet the reformulated expression, the compact matrix form in the fourth line, unlocks hidden structural features. Two striking observations follow:

\begin{itemize}
    \item The fourth line contains four zero entries, two of which are independent since $\widetilde{\mathbf{H}}'$ is constructed from $\mathbf{H}'$ by $\widetilde{\textbf{H}}'=-\epsilon \textbf{H}'^* \epsilon$. These ``empty slots'' suggest a natural possibility: they could be filled by new scalar fields, thereby accommodating additional degrees of freedom (DoFs)  in the theory.
    \item The appearance of $\mathbf{Q}_R$ as a doublet---built from $SU(2)_L$ singlets---suggests a latent algebraic structure. While $\mathbf{Q}_L$ transforms nontrivially under the electroweak gauge group, $\mathbf{Q}_R$ does not participate in weak interactions, and therefore cannot transform as an $SU(2)_L$ multiplet. Nevertheless, its doublet form suggests a new symmetry acting on $\mathbf{Q}_R$---one that is distinct from conventional Yang--Mills gauge symmetries, yet respects the multiplet organization revealed by the reformulation.
\end{itemize}

Together, these features reveal a remarkable property of the reformulated SM: despite being fully equivalent to the conventional formulation, its mathematical structure manifests an intrinsic capacity for extension, a flexibility entirely concealed in the original presentation. Consequently, if we recognize that the SM admits the reformulation shown in Eq.(\ref{807}), it becomes immediately apparent that the inherent mathematical structure of the SM explicitly reveals a natural capacity to accommodate new scalar DoFs and new symmetries. Let us now proceed to analyze this from both new scalar DoFs and new symmetry perspectives.

Regarding symmetries, our preliminary analysis (to be elaborated in Sec.~\ref{jiajia} indicates that while the doublet $\mathbf{Q}_R$ cannot transform under any $SU(2)$ representation---as required by the experimental fact that right-handed quarks do not participate in weak interactions---it may nonetheless support a different kind of gauge symmetry. Having ruled out Yang--Mills-type extensions based on non-Abelian internal symmetries, we are led to revisit an earlier notion of local symmetry that predates the modern gauge paradigm. Notably, the very notion of gauge symmetry traces back to Hermann Weyl's 1929 proposal \cite{Weyl:1929fm}, which originally concerned invariance under local rescalings of field magnitudes (now referred to as Weyl symmetry), rather than phase rotations. This raises a natural question: can Weyl's original Abelian symmetry be extended to a broader algebraic structure---one compatible with the doublet form of $\mathbf{Q}_R$ and capable of organizing new degrees of freedom in a consistent manner? Such an extension would follow the conceptual paradigm pioneered by Yang and Mills \cite{Yang:1954ek}, who generalized local phase symmetry to act on internal multiplet spaces, but pursuing a different physical realization rooted in scale rather than phase transformations.

Concerning scalar degrees of freedom in the SM, it is well-established that in the minimal SM, the Higgs field stands as the sole fundamental scalar field, responsible for electroweak symmetry breaking and mass generation via the Higgs mechanism \cite{Englert:1964et,Higgs:1964pj,Guralnik:1964eu}. However, upon incorporating Weyl symmetry into the framework, one immediately encounters a tension between Weyl symmetry and the Higgs boson: Weyl symmetry requires that physics should be independent of the system of units used to describe it \cite{Weyl:1929fm}. Consequently, any dimensional quantity necessarily changes its numerical value under such a transformation. Since a dimensional real scalar field can be transformed into a constant by a Weyl transformation, the discovery of the Higgs field \cite{Discovery} as the only scalar component in the SM implies two opposing possibilities:
\begin{itemize}
\item The SM should not be extended to accommodate Weyl symmetry otherwise the observed scalar Higgs boson would be eliminated from the physical particle spectrum leading to obvious contradiction\footnote{When the Higgs boson has not been detected for more than twenty years since its theoretical proposal, there was suspicion that the Higgs boson may not exist. It was proposed that the requirement of Weyl symmetry leads to the existence of Weyl meson which absorbs the Higgs particle \cite{Cheng:1988zx}.}; 
\item The SM should be generalized to accommodate Weyl symmetry \cite{Cheng:1988zx,Cheng:1988jy}\footnote{Interested reader can see \cite{Meissner:2006zh,Quiros:2014wda,Ghilencea:2021lpa,Ghilencea:2023sti} for further reading.}, 
then in addition to the Weyl meson, one or more degrees of freedom must be introduced to neutralize this symmetry for the purpose of protecting the existence of the Higgs field. 
\end{itemize}
While the first option offers little room for development, the second aligns precisely with the structural extensibility revealed by Eq.~\eqref{807}. Indeed, the presence of ``empty slots'' in the reformulated Yukawa sector provides a natural locus for such compensating scalars. Given that a single Weyl gauge field can fix only one degree of freedom, the minimal extension requires a second scalar field---an approach explored in various contexts \cite{Singh:2011av,Kashyap:2012kg,Tang:2019uex,Ghilencea:2023wwf}. Building on this, the present work proposes a unified perspective: the same reformulation that exposes hidden scalar slots also suggests a path toward extending Weyl symmetry itself. By elevating Weyl symmetry from an Abelian rescaling to a more general transformation acting on emergent doublets like $\mathbf{Q}_R$, we aim to introduce new degrees of freedom---not as ad hoc additions, but as inevitable consequences of the SM's latent algebraic structure.

This perspective is developed in three stages. First, in Sec.~\ref{jia}, we present the compact matrix reformulation of the SM Yukawa sector. Its principal merit lies in making the model's extensibility manifest---transforming model-building from ad hoc imposition into an intrinsic theoretical imperative. As outlined above, this extensibility operates along two axes: (1) the natural incorporation of new scalar degrees of freedom into the vacant entries of the Yukawa matrix, and (2) the potential for novel symmetries acting on emergent doublets such as $\mathbf{Q}_R$. Crucially, experimental constraints---the absence of weak interactions for right-handed fermions---exclude conventional Yang--Mills extensions, thereby motivating an alternative approach based on an extended Weyl symmetry. The construction and implications of this framework are explored in Sec.~\ref{jiajia}. Building on these foundations, Sec.~\ref{sectionc} demonstrates how the combined structure consistently accommodates a second scalar field and other new degrees of freedom, while offering multiple theoretical and phenomenological advantages. We emphasize that the entire analysis remains at the classical level; quantum aspects of the extended Weyl symmetry---including potential anomalies, unitarity issues, and renormalization challenges---are expected to be nontrivial and are reserved for future investigation.

\section{Reformulation of the SM} 
\label{jia}
\subsection{Equivalent reformulation}
At the risk of verbosity, we present the parts of the SM's Lagrangian that are relevant to our subsequent discussion as follows: 
\begin{equation}
\label{lagrangiansm}
\begin{aligned}
\mathcal{L}'=&\ \ \frac{1}{2}(\textbf{D}_{\mu}\textbf{H})^{\dagger{}}\textbf{D}^{\mu}\textbf{H}-\dfrac{1}{4}\big{(}\textbf{H}^\dagger{}\textbf{H}- v^2\big{)}^2\\
&+i\textbf{L}^{\dagger{}}\bar{\sigma}^{\mu}\mathcal{D}_{\mu}\textbf{L}+i\textbf{Q}_L^{\dagger{}}\bar{\sigma}^{\mu}\mathcal{D}_{\mu}\textbf{Q}_L\\
&+i\nu_R^{\dagger{}}\sigma^{\mu}\partial_{\mu}\nu_R+ie_R^{\dagger{}}\sigma^{\mu}D_{\mu}e_R\\
&+iu_R^{\dagger{}}\sigma^{\mu}D_{\mu}u_R+id_R^{\dagger{}}\sigma^{\mu}D_{\mu}d_R\\
&-\textbf{L}^{\dagger{}}\textbf{H}e_R-\textbf{Q}_L^{\dagger{}}\textbf{H}d_R-\textbf{Q}_L^{\dagger{}}\widetilde{\textbf{H}}u_R\\
&-e_R^{\dagger{}}\textbf{H}^{\dagger^{}}\textbf{L}-d_R^{\dagger{}}\textbf{H}^{\dagger^{}}\textbf{Q}_L-u_R^{\dagger{}}\widetilde{\textbf{H}}^{\dagger^{}}\textbf{Q}_L.
\end{aligned}
\end{equation}
Here, $\textbf{L} = \begin{pmatrix} \nu_L & e_L \end{pmatrix}^{\!T}$ is the $SU(2)_L$ doublet formed by the left-handed neutrino and electron ($\nu_L$ and $e_L$); $\nu_R$ and $e_R$ are the right-handed neutrino and electron which are also $SU(2)_L$ singlets; the covariant derivatives take the following form:
\begin{equation}
\label{1010}
\textbf{D}^{\mu}\textbf{H}=(\partial^{\mu}-iW^{\mu}+\frac{i}{2}B^{\mu})\textbf{H},
\end{equation}
\begin{equation}
\mathcal{D}^{\mu}\textbf{L}=(\partial^{\mu}-iW^{\mu}-\frac{i}{2}B^{\mu})\textbf{L},
\end{equation}
\begin{equation}
\mathcal{D}^{\mu}\textbf{Q}_L=(\partial^{\mu}-iW^{\mu}+\frac{i}{6}B^{\mu})\textbf{Q}_L,
\end{equation}
\begin{equation}
D^{\mu}e_R=(\partial^{\mu}-iB^{\mu})e_R,
\end{equation}
\begin{equation}
D^{\mu}u_R=(\partial^{\mu}+\frac{2i}{3}B^{\mu})u_R,
\end{equation}
\begin{equation}
\label{101010}
D^{\mu}d_R=(\partial^{\mu}-\frac{i}{3}B^{\mu})d_R,
\end{equation}
in which $W^{\mu}=W^{\mu}_iT^i$ ($T^i=\sigma^i /2$ in which $\sigma^i$ are the Pauli matrices) and $B^{\mu}$ are gauge fields for $SU(2)_L$ and $U(1)_Y$ symmetries respectively; furthermore, the weak hypercharges for $\textbf{L}$, $\textbf{Q}_L$ and $\textbf{H}$ are $Y_L=1$, $Y_{Q_L}=-1/3$ and $Y_H=-1$ respectively. Let us pay attention to the following three somewhat trivial facts:
\begin{itemize}
\item
The $SU(2)_L\times U(1)_Y$ transformation of the $SU(2)_L$ doublets $\textbf{L}$ and $\textbf{Q}_L$ can be expressed as
\begin{equation}
\label{trivialu1}
\begin{aligned}
\ \ \ \qquad \textbf{L}&\longrightarrow e^{i\alpha_iT^i}e^{i\theta}\textbf{L}=e^{i\alpha_iT^i}e^{2i\theta T^0}\textbf{L},\\
\textbf{Q}_L&\longrightarrow e^{i\alpha_iT^i}e^{-i\theta /3}\textbf{Q}_L=e^{i\alpha_iT^i}e^{(-2i\theta /3) T^0}\textbf{Q}_L
\end{aligned}
\end{equation}
with $T^0=\sigma^0/2$, which means that the $SU(2)_L$ doublets $\textbf{L}$ and $\textbf{Q}_L$ can also be viewed as \textit{doublets of the Abelian and unitary group generated by the generator $T^0=\sigma^0/2$}.

\item
The $SU(2)_L$ Higgs doublet $\textbf{H}$ transform under $SU(2)_L\times U(1)_Y$ transformation as $\textbf{H} \rightarrow  e^{i\alpha_i T^i}e^{-i\theta}\textbf{H}$.
Generalizing this $2\times 1$ matrix  $\textbf{H}$ to a $2\times 2$ matrix $\textbf{H}'$ with its first column matrix entries all being zero:
\begin{equation}
\label{2222}
\textbf{H}=\begin{pmatrix} \Phi_1\\ \Phi_2 \end{pmatrix}\longrightarrow \textbf{H}'=\begin{pmatrix}0& \Phi_1\\ 0& \Phi_2 \end{pmatrix},
\end{equation}
it is apparent to see that this new $2\times 2$ matrix transform under $SU(2)_L\times U(1)_Y$ transformation in the same way as that for the Higgs doublet $\textbf{H}$:
\begin{equation}
\label{1414}
\begin{pmatrix}0& \Phi_1\\ 0& \Phi_2 \end{pmatrix}\longrightarrow e^{i\alpha_i T^i}e^{-i\theta}\begin{pmatrix}0& \Phi_1\\ 0& \Phi_2 \end{pmatrix}.
\end{equation}
In other words, the second column matrix entries transform exactly the same as that for a $SU(2)_L$ doublet, and the first column matrix entries will always keep to be zero under the $SU(2)_L$ transformation.

\item
Just as the same as constructing the charge-conjugated Higgs field $\widetilde{\textbf{H}}$ from the Higgs doublet $\textbf{H}$ by $\widetilde{\textbf{H}} = \epsilon \textbf{H}^*$, we can construct a new matrix $\widetilde{\textbf{H}}'$ from $\textbf{H}'$ by $\widetilde{\textbf{H}}'=-\epsilon \textbf{H}'^* \epsilon$. It is straightforward to check that the matrix $\widetilde{\textbf{H}}'$ transform under the $SU(2)_L\times U(1)_Y$ transformation as: 
\begin{equation}
\label{1515}
\begin{pmatrix}\Phi^*_2& 0\\ -\Phi^*_1& 0 \end{pmatrix}\longrightarrow e^{i\alpha_i T^i}e^{i\theta}\begin{pmatrix}\Phi^*_2& 0\\ -\Phi^*_1& 0 \end{pmatrix}.
\end{equation}
Now it is the first column matrix entries transform as a $SU(2)_L$ doublet, while the second column matrix entries always keep to be zero under the $SU(2)_L$ transformation.
\end{itemize}

It is natural to notice that, with the $2\times 2$ matrices $\textbf{H}'$ and $\widetilde{\textbf{H}}'$ at hand, \textit{if} the right-handed spinor pairs are some kind of \textit{well-defined} doublets denoted as $\textbf{R} = \begin{pmatrix} \nu_R & e_R \end{pmatrix}^{\!T}$ and $\textbf{Q}_R = \begin{pmatrix} u_R & d_R \end{pmatrix}^{\!T}$, one has the potential to reformulate $\mathcal{L}'$ into an equivalent but more concise form, as shown in Eq.(\ref{807}).
The question is to which specific symmetry group do these doublets ($\textbf{R}$ and $\textbf{Q}_R$) belong? In the SM both left- and right-handed spinors interact via the $U(1)_Y$ symmetry. A trivial fact about the left-handed spinors is that the $U(1)_Y$ symmetry of the $SU(2)_L$ doublets ($\textbf{L}$ and $\textbf{Q}_L$) can be represented by the Abelian and unitary group generated by the generator $T^0=\sigma^0/2$ (see (\ref{trivialu1})). This trivial fact motivates a feasible possibility: \textit{$\mathbf{R}$ and $\mathbf{Q}_R$ are also doublets of a unitary and Abelian group represented by appropriate $2\times 2$ matrices}---not necessary just the same unitary and Abelian group as that for $\textbf{L}$ and $\textbf{Q}_L$ since the left- and right-handed spinors are different to each other.


The group element of the most general unitary and Abelian group represented by appropriate $2\times 2$ matrices takes the form as
\begin{equation}
\label{77}
e^{2ai\theta T^0}e^{2bi\omega_iT^i}=e^{2ai\theta T^0+2bi\omega_iT^i},
\end{equation}
in which $a$ and $b$ are two constants introduced for later convenience; furthermore, the parameters of arbitrary two group elements given by Eq.(\ref{77}), denoted as $e^{2ai\theta T^0+2bi\omega_iT^i}$ and $e^{2ai\alpha T^0+2bi\rho_jT^j}$, must satisfy the following constraint:
\begin{equation}
\rho_i=e^{\Omega}\omega_i
\end{equation}
in which $\Omega$ is a real scalar function. Only by obeying this constraint can elements given by Eq.(\ref{77}) form a Abelian group (see Eqs.(\ref{constraint})-(\ref{88}) for related details).

However, due to the following two reasons the group defined by Eq.(\ref{77}) cannot be the group we are looking for: firstly, this group would introduce five gauge fields into the theory which contradicts the well known fact that only one gauge field ($B^{\mu}$, see Formulas (\ref{1010})-(\ref{101010})) interacts with right-handed spinors; secondly, implementing the group element $e^{2ai\theta T^0+2bi\omega_iT^i}$ from the left to the right on the matrices $\textbf{H}'$ cannot remain the first column matrix entries still to be zero after the transformation (for $\widetilde{\textbf{H}}'$, it is the second column matrix entries cannot be maintained to be zero), which contradicts the constraints placed by the $SU(2)_L$ symmetry on the form of $\textbf{H}$ and $\widetilde{\textbf{H}}'$, see Formulas (\ref{1414}) and (\ref{1515}).
The way out of these difficulties is to impose further constraints on the parameters. This can be done through the following setting:
\begin{equation}
\omega_1=\omega_2=0,\ \ \ \ \ \ \omega_3=\theta.
\end{equation}
The constraint $\omega_1=\omega_2=0$ insures the first column matrix entries of $\textbf{H}'$ (as well as the second column matrix entries of $\widetilde{\textbf{H}}'$) always to be zero which meets the requirement of the $SU(2)_L$ symmetry; the constraint $\omega_3 = \theta$ ensures that only the gauge field $B^{\mu}$ is introduced, without adding any extra DoFs.

Now, it is clear that, even though the right-handed spinors are all $SU(2)_L$ singlets and transform under $U(1)_Y$ transformation with different charges, they can be arranged appropriately into two doublets, $\textbf{R}$ and $\textbf{Q}_R$, that belong to the unitary and Abelian group with its group element denoted as
\begin{equation}
\label{2121}
e^{2ai\theta T^0+2bi\theta T^3}=\begin{pmatrix} e^{i(a+b)\theta}&0\\ 0&e^{i(a-b)\theta} \end{pmatrix}.
\end{equation}
The values of the two constants, $a$ and $b$, can be determined by observing the behavior of $\textbf{R}$ and $\textbf{Q}_R$ under the group transformation: 
\begin{equation}
\label{1717}
\begin{aligned}
\textbf{R}&\longrightarrow \begin{pmatrix} e^{i(a_R+b_R)\theta}&0\\ 0&e^{i(a_R-b_R)\theta} \end{pmatrix}\textbf{R}= \begin{pmatrix} 1&0\\ 0&e^{2i\theta} \end{pmatrix}\textbf{R},\\
\textbf{Q}_R&\longrightarrow \begin{pmatrix} e^{i(a_{Q_R}+b_{Q_R})\theta}&0\\ 0&e^{i(a_{Q_R}-b_{Q_R})\theta} \end{pmatrix}\textbf{Q}_R\\
&\qquad =\begin{pmatrix} e^{-4i\theta/3}&0\\ 0&e^{2i\theta/3} \end{pmatrix}\textbf{Q}_R,
\end{aligned}
\end{equation}
which gives
\begin{equation}
\label{2828}
\begin{aligned}
a_R&=1=Y_L,\ \ \qquad \ \ \ b_R=-1;\\
a_{Q_R}&=-1/3=Y_{Q_L},\ \ \ \  b_{Q_R}=-1.
\end{aligned}
\end{equation}

The matrix $\textbf{H}'$ transforms in a more complicated form:
\begin{equation}
\label{1616}
\textbf{H}' \longrightarrow e^{i\alpha_i T^i}e^{-i\theta}\textbf{H}'\begin{pmatrix} e^{i(a_{H'}+b_{H'})\theta}&0\\ 0&e^{i(a_{H'}-b_{H'})\theta} \end{pmatrix}^{\dagger}.
\end{equation}
The values of $a_{H'}$ and $b_{H'}$ can be determined by the invariance of $\textbf{L}^{\dagger{}}\textbf{H}'\textbf{R}+\textbf{Q}_L^{\dagger{}}\textbf{H}'\textbf{Q}_R$ under transformations given by Formulas (\ref{1414}) and (\ref{1717}), which are
\begin{equation}
a_{H'}=-1=Y_H,\ \ \qquad \ \ \ b_{H'}=-1.
\end{equation}
Thus, Formula (\ref{1616}) takes the form as
\begin{equation}
\label{1919}
\textbf{H}' \longrightarrow e^{i\alpha_i T^i}e^{-i\theta}\textbf{H}'\begin{pmatrix} e^{2i\theta}&0\\ 0&1 \end{pmatrix}=e^{i\alpha_i T^i}\textbf{H}'\begin{pmatrix} e^{i\theta}&0\\ 0&e^{-i\theta}\end{pmatrix}.
\end{equation}
This result give rise to a delightful outcome: given Formula (\ref{1919}), it can be proved that the matrix $\widetilde{\textbf{H}}'=-\epsilon \textbf{H}'^*\epsilon$ transform in the \textit{same} way as that for $\textbf{H}'$:
\begin{equation}
\label{2020}
\widetilde{\textbf{H}}' \longrightarrow e^{i\alpha_i T^i}e^{i\theta}\widetilde{\textbf{H}}'\begin{pmatrix} 1&0\\ 0&e^{-2i\theta} \end{pmatrix}=e^{i\alpha_i T^i}\widetilde{\textbf{H}}'\begin{pmatrix} e^{i\theta}&0\\ 0&e^{-i\theta}\end{pmatrix}.
\end{equation}
Formulas (\ref{1919}) and (\ref{2020}) indicate that we can add these two terms ($\textbf{H}'$ and $\widetilde{\textbf{H}}'$) together since they transform in the same way. 
On the contrary, $\textbf{H}$ and $\widetilde{\textbf{H}}$ cannot be added together since they transform with different $U(1)_Y$ hypercharges.  In fact, at first glance,
$\textbf{H}'+\widetilde{\textbf{H}}'$ is also forbidden by the $U(1)_Y$ symmetry, see Formulas (\ref{1414}) and (\ref{1515}). Nevertheless, when all the related issues are settled appropriately by the guidance of symmetry, one can indeed add $\textbf{H}'$ and $\widetilde{\textbf{H}}'$ together, which is important for the equivalent reformulation of the SM.

It should be emphasized that all we have done in Formulas (\ref{trivialu1}), (\ref{1717}), (\ref{1919}) and (\ref{2020}) is reformulate what we already know into an equivalent new form. 
The symmetry content and DoFs of the SM are exactly the same before and after the reformulation. It is in this sense we say that the SM can be rewritten into an equivalent but more concise form as follows (terms that do not change their forms are indicated by ellipses):
\begin{equation}
\label{lagrangiansmequivalent}
\begin{aligned}
\mathcal{L}_{SM}=&\ \ \frac{1}{2}(\textbf{D}_{\mu}\textbf{H})^{\dagger{}}\textbf{D}^{\mu}\textbf{H}-\dfrac{1}{4}\big{(}\textbf{H}^\dagger{}\textbf{H}- v^2\big{)}^2+...\\
&+i\nu_R^{\dagger{}}\sigma^{\mu}\partial_{\mu}\nu_R+ie_R^{\dagger{}}\sigma^{\mu}D_{\mu}e_R\\
&+iu_R^{\dagger{}}\sigma^{\mu}D_{\mu}u_R+id_R^{\dagger{}}\sigma^{\mu}D_{\mu}d_R\\
&-\textbf{L}^{\dagger{}}\textbf{H}e_R-\textbf{Q}_L^{\dagger{}}\textbf{H}d_R-\textbf{Q}_L^{\dagger{}}\widetilde{\textbf{H}}u_R-h.c.\\
=&\frac{1}{2}\mathrm{Tr} \Big{(}(\mathbb{D}_{\mu}\textbf{H}')^{\dagger{}}(\mathbb{D}^{\mu}\textbf{H}')\Big{)}-\frac{1}{4}\mathrm{Tr} \Big{(}\textbf{H}'^{\dagger}\textbf{H}'-v^2\Big{)}^2+...\\
&+i\textbf{R}^{\dagger{}}\sigma^{\mu}\mathscr{D}_{\mu}\textbf{R}+i\textbf{Q}_R^{\dagger{}}\sigma^{\mu}\mathscr{D}_{\mu}\textbf{Q}_R   \\
&-\textbf{L}^{\dagger{}}\textbf{H}'\textbf{R}-\textbf{Q}_L^{\dagger{}}(\textbf{H}'+\widetilde{\textbf{H}}')\textbf{Q}_R-h.c.
\end{aligned}
\end{equation}
with the covariant derivatives take the following as
\begin{equation}
\mathbb{D}^{\mu}\textbf{H}'=(\partial^{\mu}-iW^{\mu}-\frac{i}{2}B^{\mu}\begin{pmatrix} 1&0\\ 0&-1\end{pmatrix})\textbf{H}',
\end{equation}
\begin{equation}
\begin{aligned}
\mathscr{D}^{\mu}\textbf{R}=&(\partial^{\mu}-\frac{i}{2}B^{\mu}+\frac{i}{2}B^{\mu}\begin{pmatrix} 1&0\\ 0&-1\end{pmatrix})\textbf{R}\\
=&(\partial^{\mu}-iB^{\mu}\begin{pmatrix} 0&0\\ 0&1\end{pmatrix})\textbf{R},
\end{aligned}
\end{equation}
\begin{equation}
\begin{aligned}
\mathscr{D}^{\mu}\textbf{Q}_R=&(\partial^{\mu}+\frac{i}{6}B^{\mu}+\frac{i}{2}B^{\mu}\begin{pmatrix} 1&0\\ 0&-1\end{pmatrix})\textbf{Q}_R\\
=&(\partial^{\mu}+\frac{i}{3}B^{\mu}\begin{pmatrix} 2&0\\ 0&-1\end{pmatrix})\textbf{Q}_R.
\end{aligned}
\end{equation}

\subsection{Abelian subgroup of $U(2)$}\label{112727}

The equivalent and concise reformulation of the SM in fact reveals that the $U(1)_Y$ symmetry of the SM possesses richer meaning than was originally recognized:
\begin{itemize}
\item
For $SU(2)_L$ doublets  $\textbf{L}$ and $\textbf{Q}_L$ in the SM, they are in the irreducible $U(1)_Y$ representation with its group element expressed as $e^{iY\theta}$ in which $Y$ is the hypercharge; nevertheless, in the reformulated version of the SM, they are equivalently and also trivially treated as \textit{doublets} of the unitary and Abelian group with its group element given by (see Formula (\ref{trivialu1}), notice that for $\textbf{L}$ and $\textbf{Q}_L$, the value of $Y$ should be taken as $Y_{L}$ and $Y_{Q_L}$ correspondingly)
\begin{equation}
e^{2Yi\theta T^0}=\begin{pmatrix}e^{iY\theta}& 0\\ 0& e^{iY\theta}\end{pmatrix}.
\end{equation}
Denoting this group as $U(2)_{Y, \ T^0, \ \theta}$ since it is an Abelian subgroup of $U(2)$ generated by $T^0$ with $\theta$ as the only free parameter, then the trivial $U(1)_Y$ symmetry extension for the left-handed spinors can be described sketchily as
\begin{equation}
\label{3232}
U(1)_Y\longrightarrow U(2)_{Y, \ T^0, \ \theta}.
\end{equation}

\item
For $SU(2)_L$ singlets $u_R$ and $d_R$ (as well as $\nu_R$ and $e_R$) in the SM, they transform under $U(1)_Y$ with different hypercharges; nevertheless, in the reformulated version of the SM, they are coalesced into \textit{doublets} belonging to the unitary and Abelian group whose group element is given by (see Formulas (\ref{2121})-(\ref{2828}), also notice that for $\textbf{R}$ and $\textbf{Q}_R$, the value of $Y$ should still be taken as $Y_{L}$ and $Y_{Q_L}$ correspondingly)
\begin{equation}
\label{3131}
e^{2Yi\theta T^0-2i\theta T^3}=\begin{pmatrix}e^{i(Y-1)\theta}& 0\\ 0& e^{i(Y+1)\theta}\end{pmatrix}.
\end{equation}
Denoting this group as $U(2)_{Y, \ T^0,\ T^3, \ \theta}$ since it is an Abelian subgroup of $U(2)$ generated by $T^0$ and $T^3$ with $\theta$ as the only free parameter, then the $U(1)_Y$ symmetry extension for the right-handed spinors can be described sketchily as
\begin{equation}
\label{3232}
U(1)_Y\longrightarrow U(2)_{Y, \ T^0,\ T^3, \ \theta}.
\end{equation}

\item
For the $SU(2)_L$ doublets $\textbf{H}$ and its charge-conjugated $\widetilde{\textbf{H}}=\epsilon\textbf{H}^*$  in the SM, they are bridges connecting spinors with different chirality and transforming under $U(1)_Y$ with opposite hypercharges; in the reformulated version of the SM, $\textbf{H}$ is upcast into a $2\times2$ matrix $\textbf{H}'$ by appending null elements in the first column, its charge-conjugated matrix $\widetilde{\textbf{H}}'$ is constructed by $\widetilde{\textbf{H}}'=-\epsilon\textbf{H}'^*\epsilon$. They have twofold transformation under $U(2)_{Y, \ T^0, \ \theta}$ and $U(2)_{Y, \ T^0,\ T^3, \ \theta}$ to insure the invariance of theory:
\begin{equation}
\begin{aligned}
\textbf{H}' \longrightarrow &\ e^{2Y_Hi\theta T^0}\textbf{H}'(e^{2Y_Hi\theta T^0-2i\theta T^3})^{\dagger}\\
&=\textbf{H}'\begin{pmatrix} e^{i\theta}&0\\ 0&e^{-i\theta}\end{pmatrix},\\
\widetilde{\textbf{H}}' \longrightarrow &\  e^{-2Y_Hi\theta T^0}\widetilde{\textbf{H}}'(e^{-2Y_Hi\theta T^0-2i\theta T^3})^{\dagger}\\
&=\widetilde{\textbf{H}}'\begin{pmatrix} e^{i\theta}&0\\ 0&e^{-i\theta}\end{pmatrix}.
\end{aligned}
\end{equation}
\end{itemize}

To sum up, the SM can be equivalently reformulated into a more concise form if one extends the $U(1)_Y$ symmetry appropriately: for left-handed spinors, the $U(1)_Y$ symmetry is trivially the same before and after the reformulation, but for clarity we can extend it as $U(1)_Y\rightarrow U(2)_{Y, \ T^0, \ \theta}$; for right-handed spinors, the $U(1)_Y$ symmetry group is extended nontrivially as $U(1)_Y \rightarrow U(2)_{Y, \ T^0,\ T^3, \ \theta}$ (see Formulas (\ref{3131}) and (\ref{3232})); for bridges connecting left- and right-handed spinors, $\textbf{H}'$ and $\widetilde{\textbf{H}}'$, their transformation rules involve both $U(2)_{Y, \ T^0, \ \theta}$ and  $U(2)_{Y, \ T^0,\ T^3, \ \theta}$. Therefore, from a symmetry-theoretic perspective, the reformulation of the SM alters its gauge symmetry group structure from $SU(3)_c \times SU(2)_L \times U(1)_Y$ to
\begin{equation}
SU(3)_c \times SU(2)_L \times 
\begin{cases}
U(2)_{Y, T^0, \theta} & \text{for } \mathbf{L} \text{ and } \mathbf{Q}_L; \\
U(2)_{Y, T^0, T^3, \theta} & \text{for } \mathbf{R} \text{ and } \mathbf{Q}_R.
\end{cases}
\end{equation}

Building upon these works, we are now in the position to give the transformation rules for $\textbf{L}$, $\textbf{Q}_L$, $\textbf{R}$,  and $\textbf{Q}_R$ under $U(2)_{T^0,\ \theta}$ and $U(2)_{Y, \ T^0,\ T^3, \ \theta}$ as follows:
\begin{equation}
\begin{aligned}
\label{4141}
&\textbf{L} \longrightarrow e^{i\alpha_iT^i}e^{2i\theta T^0}\textbf{L},\\
&\textbf{Q}_L\longrightarrow e^{i\alpha_iT^i}e^{-2i\theta T^0 /3}\textbf{Q}_L,
\end{aligned}
\end{equation}
\begin{equation}
\begin{aligned}
\label{4242}
&\textbf{R} \longrightarrow e^{2i\theta (T^0-T^3)}\textbf{R},\\
&\textbf{Q}_R\longrightarrow e^{2i\theta (-\frac{T^0}{3}-T^3)}\textbf{Q}_R.
\end{aligned}
\end{equation}

These transformation rules clarify an important structural feature of the reformulated Standard Model: while the left-handed multiplets $\mathbf{L}$ and $\mathbf{Q}_L$ transform under the familiar electroweak symmetry, the right-handed fields---though originally $SU(2)_L$ singlets---are now organized into doublets $\mathbf{R}$ and $\mathbf{Q}_R$. In this form, their Abelian phase transformations (Eqs.(\ref{4141})-(\ref{4242})) are naturally expressed using the doublet basis, revealing a latent algebraic structure. Crucially, this reorganization is not an ad hoc extension; it follows directly from the internal consistency of the reformulated SM. Although no new physical symmetry is present at this stage, the doublet representation of the right-handed fields provides a natural carrier space for additional symmetries beyond the original gauge group. As we will show in the next section, this structural insight enables a systematic generalization: by endowing these right-handed doublets with an extended Weyl-type symmetry---a local symmetry acting independently on the newly formed multiplets---one obtains a classical framework that formally incorporates the Standard Model at the level of the action. Thus, the reformulation is more than a notational convenience; it uncovers an intrinsic representational flexibility that both accommodates the current theory and guides its principled extension.

\section{Extended Weyl symmetry}\label{jiajia}
\subsection{Formal Definition}\label{sectiona}

The reformulation developed in Sec.~\ref{jia} goes beyond a mere field redefinition: it systematically groups the right-handed fermion singlets---originally inert under $SU(2)_L$---into doublets and recasts the Higgs doublet as a $2\times2$ matrix. Although these doublets do not transform under the electroweak $SU(2)_L$, their unified structure suggests a natural extension: a local scale symmetry acting on each multiplet as a whole. We thus introduce the \textbf{extended Weyl symmetry}, a local gauge principle that exploits the representational flexibility uncovered by the reformulation. In what follows, we give the formal definition of this symmetry, laying the groundwork for the extended theory developed in subsequent sections.

When extending Abelian $U(1)$ transformation $e^{i\theta}$ to non-Abelian $SU(N)$ transformation, one utilizes the formula
\setlength\abovedisplayskip{4pt}
\setlength\belowdisplayskip{4pt}
\begin{equation}
\label{gw1}
e^{i\theta}\longrightarrow e^{i\alpha_a T^a}
\end{equation}
with $\alpha_a$ the real and localized group parameters and $T^a$ the generators of the group satisfying
\begin{equation}
\label{commutation relation}
[T^a,T^b]=if^{abc}T^c
\end{equation}
with $f^{abc}$ the structure coefficients of the group. It seems that we can similarly promote Abelian Weyl transformation $e^{\omega}$ to a extended form as
\begin{equation}
\label{generalized Weyl transformatioFormulan1}
e^{\omega}\longrightarrow e^{\omega_a T^a},
\end{equation}
with $\omega_a$ now arbitrary real functions acting as transformation parameters and $T^a$ again the group generators the same as those for $SU(N)$. However, the commutation relation Eq.(\ref{commutation relation}) together with the well known  Baker-Campbell-Hausdorff (BCH) formula
\begin{equation}
\label{BCH}
e^Ae^B=e^{A+B+\frac{1}{2}[A,B]+\frac{1}{12}[A,[A,B]]+...}
\end{equation}
shows that two successive transformations given in Formula (\ref{generalized Weyl transformatioFormulan1}) lead to
\begin{equation}
\label{generalized Weyl transformation2}
\begin{aligned}
e^{\omega_a T^a} e^{\rho_b T^b}&=e^{\omega_a T^a+\rho_b T^b+\frac{1}{2}[\omega_a T^a,\rho_b T^b]+...}\\
&=e^{(\alpha_a +i\beta_a)T^a}
\end{aligned}
\end{equation}
with $\alpha_a$ and $\beta_a$ real functions of $\omega_a$ and $\rho_a$. That is, transformation $e^{\omega_a T^a}$ with $\omega_a$ arbitrary real functions is not closed to itself, thus it cannot form a group. Nevertheless, using the standard procedures in Yang--Mills gauge theory, one can see that transformation with complex group parameters given in Eq.(\ref{generalized Weyl transformation2}) in fact leads to the complexification of the Yang--Mills gauge theory. Thus transformation $e^{\omega_a T^a}$ with $\omega_a$ arbitrary real functions, even though cannot form a symmetry group on its own, may has potential value in studying the complexification of the Yang--Mills gauge theory. Diving into this topic would be a departure from the focus of the present work, so we leave it for future interest. 

The closure of the transformation $e^{\omega_a T^a}$ can be realized by imposing restriction on the transformation parameters. Demanding the parameters of arbitrary two transformations, denoted as $e^{\omega_a T^a}$ and $e^{\rho_b T^b}$, to satisfy the following constraint:
\begin{equation}
\label{constraint}
\rho_a=e^{\Omega} \omega_a,
\end{equation}
in which $\Omega$ is an arbitrary real-valued scalar function. The arbitrariness of $\Omega$ is reflected in the fact that different transformations are connected to each other through different real-valued functions. For instance, given three distinct transformations $e^{\alpha_a T^a}$, $e^{\beta_b T^b}$, and $e^{\gamma_c T^c}$, the constraints between $\alpha$, $\beta$, and $\gamma$ can be expressed as $\alpha_a= e^{\Omega_1} \beta_a= e^{\Omega_2} \gamma_a$, with $\Omega_1$ and $\Omega_2$ different to each other. Then, under the constraint (\ref{constraint}), the commutator of $\omega_a T^a$ and $\rho_b T^b$ is
\begin{equation}
\label{99}
[\omega_a T^a,\ \rho_b T^b]=e^{\Omega} [\omega_a T^a,\ \omega_b T^b].
\end{equation}
Since $a$ and $b$ are dummy indices, the expressions $\omega_a T^a$ and $\omega_b T^b$ are identical under the Einstein summation convention---they represent exactly the same quantity, which means that 
\begin{equation}
\label{11223344}
[\omega_a T^a,\rho_b T^b]=e^{\Omega}[\omega_a T^a, \omega_b T^b]=0.
\end{equation}
Using the BCH formula (\ref{BCH}), Eq.(\ref{11223344}) indicates that two successive transformations lead to 
\begin{equation}
\label{88}
e^{\omega_a T^a} e^{\rho_b T^b}=e^{\omega_a T^a+\rho_b T^b}=e^{(\omega_a+\rho_a) T^a}=e^{\kappa_a T^a}
\end{equation}
with $\kappa_a$ here being $N^2-1$ new real-valued functions induced by $\omega_a$ and $\rho_a$ through $\kappa_a=\omega_a+\rho_a$, showing the closure of the transformation $e^{\omega_a T^a}$. 

For clarity, let us make the following explicit statements: the set of the extended Weyl transformation $e^{\omega_a T^a}$ is defined by the following three conditions:

\begin{itemize}
\item Parameters $\omega_a$ are $N^2-1$ real-valued functions with $a=1, 2, ..., N^2-1$;
\item $T^a$ are generators for the $SU(N)$ group satisfying $[T^a,T^b]=if^{abc}T^c$;
\item For any two elements $e^{\omega_a T^a}$ and $e^{\rho_b T^b}$ in this set,
 \ parameters $\omega_a$ and $\rho_a$ satisfy the constraint $\rho_b=e^{\Omega}\omega_b$ with $\Omega$ an arbitrary real-valued function.
\end{itemize}
Based on the result given by Eq.(\ref{88}), it is straightforward to prove that this set forms a group under ordinary matrix multiplication. Furthermore, according to this definition, arbitrary non-Abelian $SU(N)$ transformation can induce a corresponding extended Weyl transformation which we denote it as $GW(N)$ transformation\footnote{The extended Weyl transformation induced by $SU(N)$ transformations is a generalization of the conventional Weyl transformation. We denote it as $GW(N)$ rather than $EW(N)$ to avoid potential terminological confusion with the standard Electroweak Theory (commonly abbreviated as EW theory).}. In this perspective, Weyl symmetry can be seen as induced by $U(1)$ symmetry, we thus denote Weyl symmetry as $W(1)$ symmetry. Naturally, when a theory is invariant under this $GW(N)$ transformation, we say that this theory has the extended Weyl symmetry (i.e., $GW(N)$ symmetry). 

It should be noticed that, even though the way we introduce the $GW(N)$ symmetry is entirely modeled after the approach of constructing non-Abelian symmetries from Abelian gauge symmetry (see Formulas (\ref{gw1}) and (\ref{generalized Weyl transformatioFormulan1})), and it needs $N^2-1$ (equals to the number of generators of $SU(N)$) independent variables to characterize this transformation, the $GW(N)$ symmetry is still a Abelian symmetry (see Formula (\ref{88}))\footnote{We are deeply indebted to the anonymous expert who compassionately alerted us to this critical issue. Given our deep-seated mindset, we could scarcely have identified independently.}. Furthermore, this construction is purely classical; quantum consistency---particularly the cancellation of potential Weyl anomalies---remains an open issue.

\subsection{Extending $SU(3)_c$}\label{su3}
The symmetry group of the SM including the Weyl symmetry is $SU(3)_c\times SU(2)_L\times \big{(}U(1)_Y\times W(1) \big{)}$ before the spontaneously symmetry breaking (SSB) of the vacuum. According to the results obtained in Sec.~\ref{sectiona}, it is straightforward to expect that both $SU(2)_L$ and $SU(3)_c$  should induce corresponding $GW(2)_L$ and $GW(3)_c$ symmetries respectively. 
It is indeed this case for $SU(3)_c$. Extending the $SU(3)_c$ symmetry to incorporate the $GW(3)_c$ symmetry means that the corresponding transformation rule for quark triplet $q_c$ changes from $q_c\longrightarrow e^{i\alpha_a T^a}q_c$ to
\begin{equation}
q_c\longrightarrow e^{i\alpha_a T^a}e^{\omega_b T^b}q_c,
\end{equation}
with $T^a$ the generators for $SU(3)$ and $GW(3)$. Despite the special constraint on the parameter $\omega_a$ (see Eq.(\ref{constraint})),  generally one will have following result:
\begin{equation}
\label{9191}
e^{i\alpha_a T^a}e^{\omega_b T^b}q_c=e^{(\beta_a +i\gamma_a)T^a}q_c
\end{equation}
with $\beta_a$ and $\gamma_a$ \textit{all} being real functions of $\alpha_a$ and $\omega_a$. Therefore, $SU(3)_c$ and $GW(3)_c$ are non-commuting and not mutually independent. The group product between them constitutes a non-trivial extension beyond mere direct ($\times$) or semidirect ($\rtimes$) products, so we denote the product of $SU(3)_c$ and $GW(3)_c$ symbolically as 
\begin{equation}
SU(3)_c\odot GW(3)_c.
\end{equation}

The covariant derivative related to this symmetry can be defined accordingly:
\begin{equation}
\begin{aligned}
D_{\mu}&=\partial_{\mu}-i(G_{\mu}^a-i\mathcal{W}_{\mu}^a)T^a\\
&=\partial_{\mu}-iG_{\mu}^aT^a-\mathcal{W}_{\mu}^aT^a,
\end{aligned}
\end{equation}
with $(G_{\mu}^a-i\mathcal{W}_{\mu}^a)T^a$ being the gauge fields related to gauge transformation $e^{(\beta_a +i\gamma_a)T^a}$. It is then easy to figure out that, there is no interaction between $(G_{\mu}^a-i\mathcal{W}_{\mu}^a)T^a$ and the Higgs field; furthermore, similar to the fact that Weyl meson has no interactions with SM particles \cite{Hayashi:1976uz}, $\mathcal{W}_{\mu}^a$ also do not couple to fermions and gauge particles in the SM. In this perspective, $(G_{\mu}^a-i\mathcal{W}_{\mu}^a)T^a$ as a whole are gauge fields related to symmetry group $SU(3)_c\odot GW(3)_c$, and in the field combination $(G_{\mu}^a-i\mathcal{W}_{\mu}^a)T^a$, $\mathcal{W}_{\mu}^a$ are potential dark matter candidates, while the $G^a_{\mu}$ fields correspond to the gluon fields in the SM. As for the purpose of introducing a second scalar field into the SM consistently and elegantly, enlarging the symmetry from $SU(3)_c$ to $SU(3)_c\odot GW(3)_c$ has no direct benefit. 

From an alternative perspective, the group product defined by Formula (\ref{9191}) in fact leads to the complexification of $SU(3)_c$ gauge theory. Correspondingly, the $SU(3)_c$ covariant derivative changing its form as
\begin{equation}
\label{cd}
\begin{aligned}
D_{\mu}&=\partial_{\mu}-iG_{\mu}^cT^c\\
\longrightarrow D_{\mu}&=\partial_{\mu}-i(G_{\mu}^c-i\mathcal{W}_{\mu}^c)T^c
\end{aligned}
\end{equation}
can be effectively recognized as the complexification of $SU(3)_c$ gauge fields. While the complexification of the strong interaction holds significant theoretical promise and warrants deeper exploration, this direction diverges from the core objectives of the present work, we therefore defer such investigations to future studies.

\subsection{Extending $SU(2)_L$} \label{2.2}

The situation for $SU(2)_L$ is more subtle. In fact naively enlarging original $SU(2)_L$ doublets in the SM to have $SU(2)_L\times GW(2)_L$ symmetry would cause difficulties for constructing a $GW(2)$ invariant Lagrangian. To extend the SM to include the $GW(2)$ symmetry consistently, some delicate preparatory work needs to be done.
 
 \subsubsection{Three settings on the field variables}

The first setting is to require the left-handed $SU(2)_L$ doublets in the SM, denoted as $\textbf{L}$ for leptons and $\textbf{Q}_L$ for quarks, to be still  $SU(2)$ doublets but at the same time be $GW(2)_R$ singlets, which means that $\textbf{L}$ and $\textbf{Q}_L$ transform under $SU(2)_L\times GW(2)_R$ transformation as
\begin{equation}
\label{n1}
\textbf{L}\longrightarrow  e^{i\alpha_i T^i}\textbf{L},
\end{equation}
\begin{equation}
\label{n2}
\textbf{Q}_L\longrightarrow  e^{i\alpha_i T^i}\textbf{Q}_L.
\end{equation}

The second setting is that, for $SU(2)_L$ singlets $u_R$ and $d_R$ that have already been coalesced into \textit{doublets} $\textbf{Q}_R$ belonging to the unitary and Abelian group $U(2)_{Y, \ T^0,\ T^3, \ \theta}$ (see Sec.~\ref{jia} for detail), they are required to be a $GW(2)_R$ doublet simultaneously. This assignment is motivated by the desire to treat the right-handed doublet $\textbf{Q}_R$ on equal algebraic footing with $\textbf{Q}_L$, enabling a parity-symmetric formulation. This $GW(2)_R$ doublet but $SU(2)_L$ singlet transform under $SU(2)_L\times GW(2)_R$  as 
\begin{equation}
\label{n4}
\textbf{Q}_R\longrightarrow e^{\omega_i T^i}\textbf{Q}_R
\end{equation}
with $T^i=\sigma^i /2$ in which $\sigma^i$ are the Pauli matrices. For the lepton sector, in the SM neutrinos are set to be massless and there is no right-handed neutrinos. However, it has been shown in Sec.~\ref{jia} that, in order to form a $U(2)_{Y, \ T^0,\ T^3, \ \theta}$ doublet $\textbf{R}$ similar to the case in the baryon sector, one has to introduce a right-handed neutrino $\nu_R$\footnote{Since there are three generations of leptons, according to the idea in the present work, in fact three right-handed neutrinos should be introduced. However, for simplicity and also not to lose the generality, we stick only to one generation. } into the theory. Here we further require the $U(2)_{Y, \ T^0,\ T^3, \ \theta}$ doublet $\textbf{R}$ to be a $GW(2)$ doublet. This $GW(2)_R$ doublet but $SU(2)_L$ singlet transforms under $SU(2)_L\times GW(2)_R$ as
\begin{equation}
\label{n3}
\textbf{R}\longrightarrow e^{-\omega_i T^i}\textbf{R},
\end{equation} 
in which the minus sign on the exponent is for later consistency. It will be shown in Sec.~\ref{sectionc} that this kind of neutrino does have very good properties one would expect a neutrino to have, thus there are good reasons for such introduction of the right-handed neutrino---or putting it in another way: the $GW(2)_R$ symmetry demands the existence of the right-handed neutrino. 

The third setting is enlarging the $SU(2)_L$ Higgs doublet $\textbf{H}=\begin{pmatrix} \Phi_1\\ \Phi_2 \end{pmatrix}$ in the SM into two $2\times2$ Higgs matrices as
\begin{equation}
\label{Higgs multiplet1}
\textbf{H}_1=\begin{pmatrix} \Phi_1&\Phi_3\\ \Phi_4&\Phi_2 \end{pmatrix},\qquad \textbf{H}_2=\begin{pmatrix} \Phi_5&\Phi_7\\ \Phi_8&\Phi_6 \end{pmatrix}.
\end{equation} 
It is required that  $\textbf{H}_1$ and $\textbf{H}_2$ both have $4$ DoFs (nevertheless $\Phi_1 \sim \Phi_8$ do not need to be real scalar fields), thus there are $8$ real DoFs in them. Furthermore, $\textbf{H}_1$ and $\textbf{H}_2$ are defined to transform under $SU(2)_L\times GW(2)_R$ respectively as
\begin{equation}
\label{higgs}
\begin{aligned}
\textbf{H}_1&\longrightarrow e^{i\alpha_i T^i}\textbf{H}_1e^{\omega_j T^j},\\
\textbf{H}_2&\longrightarrow e^{i\alpha_i T^i}\textbf{H}_2e^{-\omega_j T^j}.
\end{aligned}
\end{equation}


It is now apparent that we can use $\textbf{L}$, $\textbf{Q}_L$,  $\textbf{R}$, $\textbf{Q}_R$, $\textbf{H}_1$ and $\textbf{H}_2$ to construct $SU(2)_L\times GW(2)_R$ invariant Lagrangian.  Taking the following Yukawa-like Lagrangian term for an example:
\begin{equation}
\label{yukawa}
\begin{aligned}
\mathcal{L}_{\tiny Yukawa-like}=&\textbf{L}^{\dagger{}}\textbf{H}_1\textbf{R}+\textbf{R}^{\dagger{}}\textbf{H}_1^{\dagger{}}\textbf{L} \\
&+\textbf{Q}_L^{\dagger{}}\textbf{H}_2\textbf{Q}_R+\textbf{Q}_R^{\dagger{}}\textbf{H}_2^{\dagger{}}\textbf{Q}_L.
\end{aligned}
\end{equation}

However, such Yukawa-like term is not complete in itself since we will see in Sec.~\ref{sectionc} that this term could not give the mass of the up quark. It is also obscure how to construct other Lagrangian terms with the expected $SU(2)_L\times GW(2)_R$ symmetry. To deal with these problems systematically, we need a special tool to help.

\subsubsection{Infeld-Van der Waerden-like Notation}  \label{2.2.2}

In the simplest terms, the rules of Infeld-Van der Waerden Notation can be summarized as follows \footnote{See \cite{Muller-Kirsten:1986ysr}, \cite{Robinson:2011lia} and \cite{Schwichtenberg:2018dri} for  excellent pedagogical reviews.}:
\begin{itemize}
\item Lower-undotted index stands for left-handed spinor while upper-dotted index stands for right-handed spinor;  dotted and undotted indices turn to each other by complex conjugation;
\item Spinor indices are raised and lowered by the spinor metric denoted as $\epsilon$; one must always combine an upper with a lower index of the same type (dotted or undotted) in order to get Lorentz invariant terms.
\item The spinor metric $\epsilon$ remains invariant under Lorentz transformations; the spinor indices of $\sigma^{\mu}_{\ a\dot{b}}$ and $\bar{\sigma}_{\mu}^{\ \dot{a}b}$ transform covariantly or contravariantly according to their positions. They interconvert spacetime four-vectors and spinors through spinor-spacetime index contraction.
\end{itemize}
This notation system provides an efficient way to work with two-component spinors conveniently. Its advantage can be utilized to construct $SU(2)_L\times GW(2)_R$ invariant Lagrangian. To do this, let us notice the fact that the ingredients of $\textbf{L}$ ($\nu_{L}$ and $e_{L}$) are left-handed spinors, and the ingredients of $\textbf{R}$ ($\nu_R$ and $e_R$) are right-handed spinors. They transform under Lorentz transformation as (using $\nu_L$ and $\nu_R$ for example)
\begin{equation}
\label{n5}
\begin{aligned}
\nu_L&\longrightarrow e^{ia_i T^i+b_j T^j}\nu_L,\\
\nu_R&\longrightarrow e^{ia_i T^i-b_j T^j}\nu_R,
\end{aligned}
\end{equation}
with $a_i$ and $b_j$ parameters for rotation and boost. On the other hand, it is by construction that  $\textbf{L}$, $\textbf{Q}_L$,  $\textbf{R}$ and $\textbf{Q}_R$ transform under $SU(2)_L\times GW(2)_R$ as
\begin{equation}
\label{n6}
\begin{aligned}
\textbf{L}&\longrightarrow  e^{i\alpha_i T^i}\textbf{L},\\
\textbf{Q}_L&\longrightarrow  e^{i\alpha_i T^i}\textbf{Q}_L,\\
\textbf{R}&\longrightarrow e^{-\omega_i T^i}\textbf{R},\\
\textbf{Q}_R&\longrightarrow e^{\omega_i T^i}\textbf{Q}_R.
\end{aligned}
\end{equation}
Here $\alpha_i$ and $\omega_j$ are parameters for $SU(2)_L$ and $GW(2)_R$ transformations, furthermore, a special constraint has to be put on the parameter $\omega_j$ (see Eq.(\ref{constraint})).

Comparing Formulas (\ref{n5}) and (\ref{n6}), it can be figured out that, \textit{if} $\textbf{L}$, $\textbf{Q}_L$,  $\textbf{R}$ and $\textbf{Q}_R$ would transform under $SU(2)_L\times GW(2)_R$ as
\begin{equation}
\label{n808}
\begin{aligned}
\textbf{L}&\longrightarrow  e^{i\alpha_i T^i+\omega_j T^j}\textbf{L},\\
\textbf{Q}_L&\longrightarrow  e^{i\alpha_i T^i-\omega_j T^j}\textbf{Q}_L,\\
\textbf{R}&\longrightarrow e^{i\alpha_i T^i-\omega_j T^j}\textbf{R},\\
\textbf{Q}_R&\longrightarrow e^{i\alpha_i T^i+\omega_j T^j}\textbf{Q}_R,
\end{aligned}
\end{equation}
we could provisionally interpret them as left/right-handed Weyl spinors with $\alpha_i$ and $\omega_j$ mimicking parameters for rotation and boost equivalently. This interpretation, however, remains conjectural. Their true transformation rules are governed by Eq.(\ref{n6})---yet it is precisely this framework that allows us to establish the following identifications for $\textbf{L}$, $\textbf{Q}_L$, $\textbf{R}$, and $\textbf{Q}_R$:
\begin{itemize}
\item Treating the $SU(2)_L$ doublets $\mathbf{L}$ \textit{as} a left-handed spinor with vanishing boost response. 
\item Treating the $SU(2)_L$ doublets $\textbf{Q}_L$ \textit{as} a right-handed spinor with vanishing boost response.
\item Treating the $SU(2)_L$ doublets $\mathbf{R}$ \textit{as} a right-handed spinor with vanishing rotational responsen.
\item Treating the $SU(2)_L$ doublets $\mathbf{Q}_R$ \textit{as} a left-handed spinor with vanishing rotational response.
\end{itemize}

The merit of this treatment lies in its immediate compatibility with Infeld-Van der Waerden symbols; consequently, analogous Infeld-Van der Waerden-like indices for $\textbf{L}$, $\textbf{Q}_L$, $\textbf{R}$, and $\textbf{Q}_R$ can be systematically assigned as follows:
\begin{equation}
\label{n7}
\begin{aligned}
\textbf{L}&=\textbf{L}_{a\alpha},\\
\textbf{Q}_L&=\textbf{Q}_{La}^{\ \ \ \dot{\alpha}},
\end{aligned}
\end{equation}
\begin{equation}
\label{n77}
\begin{aligned}
\textbf{R}&=\textbf{R}^{\dot{a}\dot{A}},\\
\textbf{Q}_R&=\textbf{Q}_{R\ A}^{\ \ \dot{a}}.
\end{aligned}
\end{equation}
That is to say, there are three kinds of indices involved here:
\begin{itemize}
\item The lower-undotted $a$ and the upper-dotted $\dot{a}$ are indices for conventional left- and right-handed spinors, respectively.
\item The lower-undotted index $\alpha$ ($\alpha=1,2$ for $\nu_L, e_L$) labels $\mathbf{L}$ as an $SU(2)_L$ doublet treated as a left-handed spinor; the upper-dotted index $\dot{\alpha}$ ($\dot{\alpha}=\dot{1},\dot{2}$ for $u_L, d_L$) analogously labels $\mathbf{Q}_L$ as an $SU(2)_L$ doublet mimicking a right-handed spinor.
\item The upper-dotted index $\dot{A}$ ($\dot{A}=\dot{1},\dot{2}$ for $\nu_R, e_R$) tags $\mathbf{R}$ as a $GW(2)_R$ doublet treated as a right-handed spinor; the lower-undotted index $A$ ($A=1,2$ for $u_R, d_R$) assigns $\mathbf{Q}_R$ to a $GW(2)_R$ doublet emulating a left-handed spinor.
\end{itemize}
Conventional left- and right-handed spinors reside in Minkowski spacetime. Their spinor indices ($a$, $\dot{b}$) interface with spacetime index $\mu$ via the $\sigma^{\mu}_{\ a\dot{b}}$ and $\bar{\sigma}_{\mu}^{\ \dot{a}b}$ mappings. Similarly, the $SU(2)_L$ and $GW(2)_R$ doublets---although \textit{treated} as left- or right-handed spinors (see Formulas (\ref{n7}) and (\ref{n77}))---exclusively inhabit the internal symmetry space. Denoting the vector index in this internal symmetry space as $M$, then the indices ($\alpha$, $\dot{\beta}$, $A$, and $\dot{B}$) interface with $M$ via the $\sigma^{M}_{\ \alpha\dot{\beta}}$, $\bar{\sigma}_{M}^{\ \dot{\alpha}\beta}$, $\sigma^{M}_{\ A\dot{B}}$, and $\bar{\sigma}_{M}^{\ \dot{A}B}$ mappings. The reason for introducing two distinct sets of indices, ($\alpha$, $\dot{\beta}$) and ($A$, $\dot{B}$), within the internal symmetry space arises from the following: while $\textbf{L}$ and $\textbf{Q}_L$ are labeled as left- and right-handed spinors in this internal space, they transform exclusively under the $SU(2)_L$ representation and remain invariant under $GW(2)_R$ actions; conversely, $\textbf{Q}_R$ and $\textbf{R}$---though analogously labeled as left- and right-handed spinors---are governed solely by $GW(2)_R$ transformations and unaffected by $SU(2)_L$ transformations. Critically, despite cohabiting the internal symmetry space, these spinor pairs are mutually decoupled: their transformation algebras operate in disjoint subspaces. Therefore, corresponding to the three types of indices, there are three types of spinor metrics and sigma matrices:

1. spinor type
\begin{equation}
\label{nn7}
\begin{cases}
\epsilon^{ab}=\epsilon^{\dot{a}\dot{b}}=-\epsilon_{ab}=-\epsilon_{\dot{a}\dot{b}}=\begin{pmatrix} 0&1\\ -1&0 \end{pmatrix},\\
\sigma^{\mu}=\sigma^{\mu}_{\ a\dot{b}},\ \ \bar{\sigma}_{\mu}=\bar{\sigma}_{\mu}^{\ \dot{a}b}.
\end{cases}
\end{equation}

2. $SU(2)_L$ type
\begin{equation}
\label{nnn7}
\ \ \begin{cases}
\epsilon^{\alpha\beta}=\epsilon^{\dot{\alpha}\dot{\beta}}=-\epsilon_{\alpha\beta}=-\epsilon_{\dot{\alpha}\dot{\beta}}=\begin{pmatrix} 0&1\\ -1&0 \end{pmatrix},\\
\sigma^{M}=\sigma^{M}_{\ \ \alpha\dot{\beta}},\ \ \bar{\sigma}_{M}=\bar{\sigma}_{M}^{\ \ \dot{\alpha}\beta}.
\end{cases}
\end{equation}

3. $GW(2)_R$ type
\begin{equation}
\label{nnnn7}
\ \ \begin{cases}
\epsilon^{AB}=\epsilon^{\dot{A}\dot{B}}=-\epsilon_{AB}=-\epsilon_{\dot{A}\dot{B}}=\begin{pmatrix} 0&1\\ -1&0 \end{pmatrix},\\
\sigma^{M}=\sigma^{M}_{\ A\dot{B}},\ \ \bar{\sigma}_{M}=\bar{\sigma}_{M}^{\ \ \dot{A}B}.
\end{cases}
\end{equation}
Based on the aforementioned rationale, it can be noticed that, $\epsilon$ and $\sigma$ given by Formula (\ref{nnn7}) should be invariant under the action of $GW(2)_R$; similarly, $\epsilon$ and $\sigma$ given by Formula (\ref{nnnn7}) also are invariant under the action of $SU(2)_L$.

By observing Eq.(\ref{higgs}), (\ref{yukawa}), (\ref{n6}), (\ref{n7}) and (\ref{n77}) carefully, the index structure of the Higgs matrices $\textbf{H}_1$ and $\textbf{H}_2$ can also be arranged as follows:
\begin{equation}
\label{n9}
\textbf{H}_1=\textbf{H}^{\ \dot{\alpha}}_{1\ \dot{A}},\quad \textbf{H}_2=\textbf{H}_{2 \alpha}^{\ \ A}.
\end{equation}
Clearly, the Higgs matrices $\textbf{H}_1$ and $\textbf{H}_2$ have $SU(2)_L$ and $GW(2)_R$ indices ($\dot{\alpha}$ and $\dot{A}$ can be turned into $\alpha$ and $A$ by complex conjugation) but no spinor index (i.e., dotted or undotted index $a$), just the same as the situation in the SM if $GW(2)_R$ symmetry is turned off.

Reminding the fact that, in the SM, the charge-conjugated Higgs field $\widetilde{\textbf{H}}$ can be constructed from the Higgs doublet $\textbf{H}$ as $\widetilde{\textbf{H}}=\epsilon \textbf{H}^*$. This field plays an important role in giving the mass of the up quark.
Based on $\textbf{H}_1$ and $\textbf{H}_2$, two new charged-conjugated matrices can also be constructed in a similar way:
\begin{equation}
\label{htilde}
\begin{aligned}
\widetilde{\textbf{H}}_{1 \alpha}^{\ \ A}&=  -\epsilon_{\alpha \beta}\big{(}\textbf{H}^{\ \dot{\beta}}_{1\ \dot{B}}\big{)}^* \epsilon^{BA},\\
\widetilde{\textbf{H}}^{\ \dot{\alpha}}_{2\ \dot{A}}&=  -\epsilon^{\dot{\alpha}\dot{\beta}}\big{(}\textbf{H}_{2 \beta}^{\ \ B}\big{)} ^*\epsilon_{\dot{B}\dot{A}}.
\end{aligned}
\end{equation}
Additionally, due to the extension of the symmetry group from $SU(2)_L$ to $SU(2)_L\times GW(2)_R$, two more matrices can be induced from $\textbf{H}_1$ and $\textbf{H}_2$:
\begin{equation}
\label{hbar}
\begin{aligned}
\bar{\textbf{H}}_{1 \alpha}^{\ \ A}&= \frac{1}{2} \sigma^{M}_{\ \ \alpha \dot{\beta}}\textbf{H}^{\ \dot{\beta}}_{1\ \dot{B}} \bar{\sigma}_{M}^{\ \ \dot{B}A},\\
\bar{\textbf{H}}^{\ \dot{\alpha}}_{2\ \dot{A}}&=\frac{1}{2}  \bar{\sigma}_{M}^{\ \ \dot{\alpha}\beta}\textbf{H}_{2 \beta}^{\ \ B} \sigma^{M}_{\ \ B \dot{A}}.
\end{aligned}
\end{equation}
It will be shown in Sec.~\ref{sectionc} that the two Higgs matrices and four induced Higgs matrices,  listed in Eqs.(\ref{n9}) $\sim$ (\ref{hbar}), act important role in the mass generation of  fermions.

In the perspective of the notation system which is developed for treating  $SU(2)_L$ and $GW(2)_R$ doublets systematically in Sec.~\ref{2.2.2}, Eqs.(\ref{n7}), (\ref{n77}), (\ref{4141}), and (\ref{4242}) indicate that, while $SU(2)_L$ indices (dotted or undotted lowercase Greek lettes, see Eq.(\ref{n7})) transform according to $U(2)_{Y, \ T^0, \ \theta}$, $GW(2)_R$ indices (dotted or undotted capital Latin letters, see Eq.(\ref{n77})) transform according to $U(2)_{Y, \ T^0,\ T^3, \ \theta}$. Then it is natural to realize that the Higgs matrices $\textbf{H}_1$ and $\textbf{H}_2$, with both $SU(2)_L$ and $GW(2)_R$ indices (see Eq.(\ref{n9})), should have twofold transformation under $U(2)_{Y, \ T^0, \ \theta}$ and $U(2)_{Y, \ T^0,\ T^3, \ \theta}$:
\begin{equation}
\label{7272}
\begin{aligned}
\textbf{H}_1&\longrightarrow e^{-2i\theta T^0} \textbf{H}_1  (e^{-2i\theta T^0-2i\theta T^3})^{\dagger}=\textbf{H}_1e^{2i\theta T^3},\\
\textbf{H}_2&\longrightarrow e^{-2i\theta T^0} \textbf{H}_2  (e^{-2i\theta T^0-2i\theta T^3})^{\dagger}=\textbf{H}_2e^{2i\theta T^3}.
\end{aligned}
\end{equation}

As a little epilogue of Sec.~\ref{2.2}, it should be noticed that, 
while $U(2)_{Y, \ T^0, \ \theta}$ commutes with $SU(2)_L$, the $U(2)_{Y, \ T^0,\ T^3, \ \theta}$ does not commute with the $GW(2)_R$. This is not the same as the case in the SM that $U(1)_Y$ commutes with all other symmetry groups. Under the joint action of $GW(2)_R$ and $U(2)_{Y, \ T^0,\ T^3, \ \theta}$, $\textbf{R}$, $\textbf{Q}_R$, $\textbf{H}_1$ and $\textbf{H}_2$ transform as
\begin{equation}
\label{6161}
\begin{aligned}
\textbf{R}\longrightarrow &e^{-\omega_i T^i}e^{2i\theta (T^0-T^3)}\textbf{R}\\
&=e^{(\eta_i(\theta, \ \omega)+i\tau_i(\theta, \ \omega))T^i}e^{2i\theta T^0}\textbf{R},\\
\textbf{Q}_R\longrightarrow &e^{\omega_i T^i}e^{2i\theta (-\frac{T^0}{3}-T^3)}\textbf{Q}_R\\
&=e^{(\gamma_i(\theta, \ \omega)+i\delta_i(\theta, \ \omega))T^i}e^{-2i\theta /3}\textbf{Q}_R,\\
\textbf{H}_1\longrightarrow &\textbf{H}_1e^{2i\theta T^3}e^{\omega_i T^i}\\
&=\textbf{H}_1e^{-(\eta_i(\theta, \ \omega)+i\tau_i(\theta, \ \omega))T^i},\\
\textbf{H}_2\longrightarrow &\textbf{H}_2e^{2i\theta T^3}e^{-\omega_i T^i}\\
&=\textbf{H}_2e^{-(\gamma_i(\theta, \ \omega)+i\delta_i(\theta, \ \omega))T^i},
\end{aligned}
\end{equation}
with $\eta$, $\tau$, $\gamma$, and $\delta$ all being real functions of $\omega_i$ and $\theta$. Therefore, similar to the product between $SU(3)_c$ and $GW(3)_c$ (see Sec.~\ref{su3}), the group product between $GW(2)_R$ and $U(2)_{Y, \ T^0,\ T^3, \ \theta}$ is another non-trivial extension beyond mere direct ($\times$) or semidirect ($\rtimes$) products, we thus denote the product between these two group as 
\begin{equation}
GW(2)_R\bigodot U(2)_{Y, \ T^0,\ T^3, \ \theta}.
\end{equation}
It is important to note that while the variables $\eta_i$ and $\tau_i$ (as well as $\gamma_i$ and $\delta_i$) may appear to constitute six parameters at first glance, they are in fact only four independent DoFs. The non-trivial structure of the $GW(2)_R\bigodot U(2)_{Y, \ T^0,\ T^3, \ \theta}$ symmetry group renders the precise definition of the associated covariant derivative a challenging task. Nevertheless, it will be explained in Sec.~\ref{sectionc} that, while the effects of the SSB of the vacuum are fully incorporated into the analysis, the apparent theoretical challenges can be satisfactorily resolved.

\enlargethispage{\baselineskip}
\subsection{The Lagrangian} \label{3.4}

Based on previous preparations, the $SU(2)_L\times U(1)_Y$ sector of the SM can be extended to accommodate $GW(2)_R$ and $W(1)$ symmetries appropriately. 
The symmetry structure of the theory is extended along the following line:
\begin{equation}
\begin{aligned}
&SU(2)_L \times U(1)_Y \\
&\rightarrow SU(2)_L \times   \left\{ \begin{aligned}
&U(2)_{Y, T^0, \theta} \ \ \ \ \ \text{for } \mathbf{L} \text{ and } \mathbf{Q}_L \\
&U(2)_{Y, T^0, T^3, \theta}\ \  \text{for } \mathbf{R} \text{ and } \mathbf{Q}_R
\end{aligned}\right\}\\
&\rightarrow SU(2)_L\times \left\{ \begin{aligned}
 &U(2)_{Y, T^0, \theta} \qquad \ \ \ \ \  \text{for } \mathbf{L} \text{ and } \mathbf{Q}_L \\
GW(2)_R &\bigodot U(2)_{Y, T^0, T^3, \theta}\ \   \text{for } \mathbf{R} \text{ and } \mathbf{Q}_R
\end{aligned}\right\} \\
&\qquad\qquad\quad \times W(1).
\end{aligned}
\end{equation}
The corresponding Lagrangian  for $\textbf{L}$, $\textbf{R}$, $\textbf{Q}_L$, $\textbf{Q}_R$, $\textbf{H}_1$, $\textbf{H}_2$ and  interactions between them can be constructed as follows:
\begin{equation}
\label{lagrangian}
\begin{aligned}
\mathcal{L}=&-\frac{1}{2}\mathrm{Tr}(\textbf{W}_{\mu\nu}\textbf{W}^{\mu\nu})-\frac{1}{4}B_{\mu\nu}B^{\mu\nu}-\frac{1}{2}\mathrm{Tr}(\mathcal{W}_{\mu\nu}\mathcal{W}^{\mu\nu})\\
&-\frac{1}{4}\Pi_{\mu\nu}\Pi^{\mu\nu}+i\textbf{L}^{\dagger{}}\textbf{I}\bar{\sigma}^{\mu}\mathcal{D}_{\mu}\textbf{L}+i\textbf{Q}_L^{\dagger{}}\textbf{I}\bar{\sigma}^{\mu}\mathcal{D}_{\mu}\textbf{Q}_L\\
&+i\textbf{R}^{\dagger{}}\mathcal{I}\sigma^{\mu}\mathscr{D}_{\mu}\textbf{R}+i\textbf{Q}_R^{\dagger{}}\mathcal{I}\sigma^{\mu}\mathscr{D}_{\mu}\textbf{Q}_R   \\
&+\frac{1}{8}\mathrm{Tr} \Big{(}(\mathbb{D}_{\mu}\bar{\textbf{H}}_1)^{\dagger{}}(\mathbb{D}^{\mu}\textbf{H}_1)+(\mathbb{D}_{\mu}\bar{\textbf{H}}_2)^{\dagger{}}(\mathbb{D}^{\mu}\textbf{H}_2)+h.c.\Big{)}\\
&-\frac{1}{2}\Big{(}\textbf{L}^{\dagger{}}(\textbf{H}_1+\bar{\textbf{H}}_2+\widetilde{\textbf{H}}_2)\textbf{R}+h.c.\Big{)} \\
&-\frac{1}{2}\Big{(}\textbf{Q}_L^{\dagger{}}(\textbf{H}_2+\bar{\textbf{H}}_1+\widetilde{\textbf{H}}_1)\textbf{Q}_R+h.c.\Big{)} \\
&-\frac{1}{8}\mathrm{Tr}\Big{(}\textbf{H}_1^\dagger{}\textbf{H}_2\textbf{H}_1^\dagger{}\textbf{H}_2+h.c.\Big{)},
\end{aligned}
\end{equation}
in which $W^{\mu}=W^{\mu}_iT^i$, $\mathcal{W}^{\mu}=\mathcal{W}^{\mu}_iT^i$, $B^{\mu}$  and $\Pi^{\mu}$ are gauge fields for $SU(2)_L$, $GW(2)_R$, $U(1)_Y$ and $W(1)$ symmetries respectively; $\textbf{W}^{\mu\nu}$, $\mathcal{W}^{\mu\nu}$, $B^{\mu\nu}$ and $\Pi^{\mu\nu}$ are the corresponding fields strengths. The covariant derivatives take the following form\footnote{Due to the non-commutativity between the $U(2)_{Y, \ T^0,\ T^3, \ \theta}$ group and the $GW(2)_R$ group, the precise formal expressions of the relative covariant derivatives are not so straightforward as listed above. Nevertheless, as it will be explained in Sec.~\ref{sectionc}, because of the SSB of the vacuum the relative covariant derivatives can indeed be set in this way.}:
\begin{align}
\mathcal{D}^{\mu}\textbf{L}=(\partial^{\mu}-iW^{\mu}-\frac{i}{2}B^{\mu})\textbf{L},\label{d1}
\end{align}
\begin{align}
\mathcal{D}^{\mu}\textbf{Q}_L=(\partial^{\mu}-iW^{\mu}+\frac{i}{6}B^{\mu})\textbf{Q}_L,\label{dql}
\end{align}
\begin{equation}
\begin{aligned}
\mathscr{D}^{\mu}\textbf{R}&=(\partial^{\mu}+\mathcal{W}^{\mu}-\frac{i}{2}B^{\mu}+\frac{i}{2}B^{\mu}\begin{pmatrix} 1&0\\ 0&-1\end{pmatrix})\textbf{R}\\
&=(\partial^{\mu}+\mathcal{W}^{\mu}-iB^{\mu}\begin{pmatrix} 0&0\\ 0&1\end{pmatrix})\textbf{R},\label{d2}
\end{aligned}
\end{equation}
\begin{equation}
\begin{aligned}
\mathscr{D}^{\mu}\textbf{Q}_R&=(\partial^{\mu}-\mathcal{W}^{\mu}+\frac{i}{6}B^{\mu}+\frac{i}{2}B^{\mu}\begin{pmatrix} 1&0\\ 0&-1\end{pmatrix})\textbf{Q}_R\\
&=(\partial^{\mu}-\mathcal{W}^{\mu}+\frac{i}{3}B^{\mu}\begin{pmatrix} 2&0\\ 0&-1\end{pmatrix})\textbf{Q}_R,\label{d3}
\end{aligned}
\end{equation}
\begin{align}
&\mathbb{D}^{\mu}\textbf{H}_1=(\partial^{\mu}-iW^{\mu}-\mathcal{W}^{\mu}-\frac{i}{2}B^{\mu}\begin{pmatrix} 1&0\\ 0&-1\end{pmatrix}-\Pi_{\mu})\textbf{H}_1,\label{d4}\\
&\mathbb{D}^{\mu}\bar{\textbf{H}}_1=(\partial^{\mu}-iW^{\mu}+\mathcal{W}^{\mu}-\frac{i}{2}B^{\mu}\begin{pmatrix} 1&0\\ 0&-1\end{pmatrix}-\Pi_{\mu})\bar{\textbf{H}}_1,\label{d5}
\end{align}
\begin{align}
\mathbb{D}^{\mu}\textbf{H}_2=(\partial^{\mu}-iW^{\mu}+\mathcal{W}^{\mu}-\frac{i}{2}B^{\mu}\begin{pmatrix} 1&0\\ 0&-1\end{pmatrix}-\Pi_{\mu})\textbf{H}_2,\label{d6}
\end{align}
\begin{align}
\mathbb{D}^{\mu}\bar{\textbf{H}}_2=(\partial^{\mu}-iW^{\mu}-\mathcal{W}^{\mu}-\frac{i}{2}B^{\mu}\begin{pmatrix} 1&0\\ 0&-1\end{pmatrix}-\Pi_{\mu})\bar{\textbf{H}}_2.\label{d7}
\end{align}

With the help of the index tools developed in Sec.~\ref{2.2.2}, the invariance of the Lagrangian (\ref{lagrangian}) under $SU(2)_L\times GW(2)_R$ transformation can be shown clearly. Taking the following three terms for example:
\begin{equation}
\mathrm{Tr}(\mathbb{D}_{\mu}\bar{\textbf{H}}_1)^{\dagger{}}(\mathbb{D}^{\mu}\textbf{H}_1)=(\mathbb{D}_{\mu}\bar{\textbf{H}}_1)^{\dagger{}\dot{A}}_{\ \ \ \dot{\alpha}}(\mathbb{D}^{\mu}\textbf{H}_1)^{\dot{\alpha}}_{\ \dot{A}};
\end{equation}
\begin{equation}
\textbf{L}^{\dagger{}}(\textbf{H}_1+\bar{\textbf{H}}_2+\widetilde{\textbf{H}}_2)\textbf{R}=\textbf{L}_{\dot{\alpha}}^{\dagger{}}(\textbf{H}_1+\bar{\textbf{H}}_2+\widetilde{\textbf{H}}_2)^{\dot{\alpha}}_{\ \dot{A}}\textbf{R}^{\dot{A}};
\end{equation}
\begin{equation}
\mathrm{Tr}(\textbf{H}_1^\dagger{}\textbf{H}_2\textbf{H}_1^\dagger{}\textbf{H}_2)=\textbf{H}_{1A}^{\dagger{}\ \ \alpha}\textbf{H}_{2\alpha}^{\ \ B}\textbf{H}_{1B}^{\dagger{}\ \ \beta}\textbf{H}_{2\beta}^{\ \ A}.
\end{equation}
All the indices are contracted in the right way thus these terms are invariant under $SU(2)_L\times GW(2)_R$ transformation. The $GW(2)_R\bigodot U(2)_{T^0, T^3, \theta}$ symmetry of this Lagrangian can be checked according to Formula (\ref{6161}), and the invariance of the other terms in the Lagrangian (\ref{lagrangian}) can be shown in the same method. 

The invariance of the kinetic terms for spinors, which is
\begin{equation}
\begin{aligned}
\mathcal{L}_{Spinors}=&i\textbf{L}^{\dagger{}}\textbf{I}\bar{\sigma}^{\mu}\mathcal{D}_{\mu}\textbf{L}+i\textbf{Q}_L^{\dagger{}}\textbf{I}\bar{\sigma}^{\mu}\mathcal{D}_{\mu}\textbf{Q}_L\\
&+i\textbf{R}^{\dagger{}}\mathcal{I}\sigma^{\mu}\mathscr{D}_{\mu}\textbf{R}+i\textbf{Q}_R^{\dagger{}}\mathcal{I}\sigma^{\mu}\mathscr{D}_{\mu}\textbf{Q}_R,
\end{aligned}
\end{equation}
may need a few more words on it. Here one can see that two matrices, $\textbf{I}$ and $\mathcal{I}$, have been introduced into the kinetic terms for the $SU(2)_L$ and $GW(2)_R$ doublets. $\textbf{I}$ is a unit matrix that carry $SU(2)_L$ indices in two ways:
\begin{equation}
\textbf{I}=\textbf{I}_{\alpha\dot{\beta}}, \ \ or \ \ \textbf{I}=\textbf{I}^{\dot{\alpha}\beta}.
\end{equation}
According to the index structure carried by them, both of them are invariant under $SU(2)_L$ transformation:
\begin{align}
\textbf{I}\longrightarrow \textbf{I}'= e^{i\alpha_iT^i}\textbf{I}e^{-i\alpha_jT^j}=\textbf{I}.
\end{align}
Since $\textbf{I}^{\dot{\alpha}\beta}$ and $\textbf{I}_{\alpha\dot{\beta}}$ are unit matrices and invariant under the $SU(2)_L$ transformation, one can safely ignore their trivial existence in the kinetic terms for $\textbf{L}$ and $\textbf{Q}_L$:
\begin{equation}
\mathcal{L}_{\textbf{L}+\textbf{Q}_L}=i\textbf{L}^{\dagger{}}\bar{\sigma}^{\mu}\mathcal{D}_{\mu}\textbf{L}+i\textbf{Q}_L^{\dagger{}}\bar{\sigma}^{\mu}\mathcal{D}_{\mu}\textbf{Q}_L,
\end{equation}
which is exactly the situation in the SM. In other words, in the SM there is no need to introduce index structure and thus needs not to think about \textit{the contraction of the $SU(2)$ indices} at all. However, the $SU(2)$ invariants do have internal index structure (see Sec.~\ref{2.2.2} for more details). This structure can be shown explicitly with the help of two unit matrices $\textbf{I}^{\dot{\alpha}\beta}$ and $\textbf{I}_{\alpha\dot{\beta}}$:
\begin{align}
&i\textbf{L}^{\dagger{}}\textbf{I}\bar{\sigma}^{\mu}\mathcal{D}_{\mu}\textbf{L}=i\textbf{L}^{\dagger{}}_{\ \dot{\alpha}}\textbf{I}^{\dot{\alpha}\beta}\bar{\sigma}^{\mu}\mathcal{D}_{\mu}\textbf{L}_{\beta},\\
&i\textbf{Q}_L^{\dagger{}}\textbf{I}\bar{\sigma}^{\mu}\mathcal{D}_{\mu}\textbf{Q}_L=i\textbf{Q}_L^{\dagger{}\alpha}\textbf{I}_{\alpha\dot{\beta}}\bar{\sigma}^{\mu}\mathcal{D}_{\mu}\textbf{Q}_L^{\ \dot{\beta}}.
\end{align}
Therefore, writing $\textbf{I}^{\dot{\alpha}\beta}$ and $\textbf{I}_{\alpha\dot{\beta}}$ evidently in the kinetic therms for $\textbf{L}$ and $\textbf{Q}_L$ can make the $SU(2)_L$ invariance as well as the expected contraction of the indices clearly. But more importantly, the inclusion of $\textbf{I}^{\dot{\alpha}\beta}$ and $\textbf{I}_{\alpha\dot{\beta}}$ in the kinetic terms for $\textbf{L}$ and $\textbf{Q}_L$ naturally inspires the introducing of the nontrivial \textit{unit-like} matrix $\mathcal{I}$ into the kinetic terms for the $GW(2)_R$ doublets $\textbf{R}$ and $\textbf{Q}_R$.  Here $\mathcal{I}$ is a matrix that carries $GW(2)_R$ indices in two ways:
\begin{equation}
\mathcal{I}=\mathcal{I}_{A\dot{B}}, \ \ or \ \ \mathcal{I}=\mathcal{I}^{\dot{A}B}.
\end{equation}
They transform under $GW(2)_R$ transformation as:
\begin{equation}
\mathcal{I}_{A\dot{B}}\longrightarrow \mathcal{I}'_{A\dot{B}}= e^{\phi_iT^i}\mathcal{I}_{A\dot{B}}e^{\phi_jT^j},
\end{equation}
\begin{equation}
\mathcal{I}^{\dot{A}B}\longrightarrow \mathcal{I}'^{\dot{A}B}= e^{-\phi_iT^i}\mathcal{I}^{\dot{A}B}e^{-\phi_jT^j}.
\end{equation}
Then it is straightforward to check that the kinetic terms for $\textbf{R}$ and $\textbf{Q}_R$ are $GW(2)_R$ invariant:
\begin{align}
&i\textbf{R}^{\dagger{}}\mathcal{I}\sigma^{\mu}\mathscr{D}_{\mu}\textbf{R}=i\textbf{R}^{\dagger{}A}\mathcal{I}_{A\dot{B}}\sigma^{\mu}\mathscr{D}_{\mu}\textbf{R}^{\dot{B}},\\
&i\textbf{Q}_{R}^{\dagger{}}\mathcal{I}\sigma^{\mu}\mathscr{D}_{\mu}\textbf{Q}_R=i\textbf{Q}_{R\dot{A}}^{\dagger{}}\mathcal{I}^{\dot{A}B}\sigma^{\mu}\mathscr{D}_{\mu}\textbf{Q}_{RB}.
\end{align}
It is also because of this $GW(2)_R$ invariance, even though $\mathcal{I}$ does transform under $GW(2)_R$ transformation, one can still \textit{treat} it \textit{as} a unit matrix in any given gauge, that is why one calls $\mathcal{I}$ a \textit{unit-like} matrix. 

Examining the Lagrangian (\ref{lagrangian}) closely, three distinctive features of it may deserve further attention.
\begin{itemize}

\item \textit{Higgs potential is given entirely by symmetry}: In the SM, the form of the Higgs potential contains ingredients that needs to be set by hand. Concretely, in the SM, the $SU(2)_L$ symmetry can only constrain the Higgs potential to have the form as
\begin{equation}
V(\textbf{H})=\dfrac{1}{4}\big{(}\textbf{H}^\dagger{}\textbf{H}\pm v^2\big{)}^2,
\end{equation}
with both plus and minus signs in the front of $v^2$ satisfying the requirements of $SU(2)_L$ symmetry. Then, \textit{in order to} spontaneously break the $SU(2)_L$ symmetry of the vacuum, the sign before $v^2$ is set \textit{by hand} to be negative. Nevertheless, in Lagrangian (\ref{lagrangian}), the Higgs potential 
\begin{equation}
\quad \quad V(\textbf{H}_1, \textbf{H}_2)=\dfrac{1}{8}\mathrm{Tr}\Big{(}\textbf{H}_1^\dagger{}\textbf{H}_2\textbf{H}_1^\dagger{}\textbf{H}_2+h.c.\Big{)}
\end{equation}
 is invariant under $SU(2)_L\times GW(2)_R$ transformation.  No dimensional constant such as $\pm v^2$ appears in the Higgs potential. In fact  any dimensional constants in the Lagrangian would violates the Weyl symmetry thus their appearance in the theory are forbidden. Therefore it is fair to say that the form of the Higgs potential in Lagrangian (\ref{lagrangian}) is constrained completely by symmetry and nothing here needs to be set by hand as that in the SM. So far it is an unanswered question that whether such Higgs potential can lead to the spontaneous symmetry breaking (SSB) of the vacuum and then gives masses of all the related fundamental particles. We will come back to this in Sec.~\ref{sectionc}.

\item \textit{Dark matter candidate}: Both the covariant derivatives of $\mathscr{D}^{\mu}$ and $\mathbb{D}^{\mu}$ include gauge fields $\mathcal{W}^{\mu}$ for $GW(2)_R$ symmetry. $\mathcal{W}^{\mu}$ does couple to the Higgs matrices $\textbf{H}_1$ and $\textbf{H}_2$. Nevertheless, similar to the case that Weyl meson does not couple to fermions and gauge particles in the SM,  it is straightforward to show that $\mathcal{W}^{\mu}$ does not couple to the $GW(2)_R$ doublets ($\textbf{R}$ and $\textbf{Q}_R$) as well as other gauge fields in the SM. Thus $\mathcal{W}^{\mu}$ can be another potential candidate for dark matter as the same as Weyl meson and $\mathcal{W}_{\mu}^c$ which is introduced in Sec.~\ref{su3}. Future precision measurements of Higgs properties may give some clues on its existence of $\mathcal{W}^{\mu}$. Even though $\mathcal{W}^{\mu}$ does not couple to the $GW(2)_R$ doublets \textbf{R} and $\textbf{Q}_R$, we still include them obviously in Eqs.(\ref{d2}) $\sim$ (\ref{d3}) for the immediate convenience.

\item \textit{Parity symmetry}: 
Parity transformation transforms the $SU(2)_L$ doublets $\textbf{L}$  and $\textbf{Q}_L$ into the $GW(2)_R$ doublets $\textbf{R}$ and $\textbf{Q}_R$ respectively (and vice versa), that is:
\begin{equation}
\label{pt1}
P: \quad \textbf{L}_{\alpha}\longleftrightarrow\textbf{R}^{\dot{A}}, \quad\quad \textbf{Q}_{L}^{\ \dot{\alpha}}\longleftrightarrow \textbf{Q}_{RA}. 
\end{equation}
From the perspective of the notation tools developed in Sec.~\ref{2.2.2} for the $SU(2)_L$ and $GW(2)_R$ doublets, Formula (\ref{pt1}) indicates the following transformation rules for the $SU(2)_L$ and $GW(2)_R$ indices:
\begin{equation}
\label{p1}
\begin{aligned}
P: \quad  &_{\alpha}\longleftrightarrow ^{\dot{A}},\quad \quad _{\dot{\alpha}}\longleftrightarrow ^A,\\
&^{\alpha}\longleftrightarrow _{\dot{A}},\quad \quad ^{\dot{\alpha}}\longleftrightarrow _A.
\end{aligned}
\end{equation}
Then the $SU(2)_L$ type and $GW(2)_R$ type sigma matrices  transforming under parity transformation can be induced:
\begin{equation}
P: \  \sigma_{M}^{\ \ \dot{A}B}\longleftrightarrow \bar{\sigma}_{M \alpha\dot{\beta}}, \ \sigma_{M}^{\ \dot{\alpha}\beta}\longleftrightarrow \bar{\sigma}_{M A\dot{B}}.  \label{pt2}
\end{equation}
These parity transformation rules indicate that the Higgs matrices $\textbf{H}_1$ and $\textbf{H}_2$, as well as matrices induced from them ( $\widetilde{\textbf{H}}_1$, $\widetilde{\textbf{H}}_2$, $\bar{\textbf{H}}_1$ and $\bar{\textbf{H}}_2$),  transform under parity transformation accordingly as:
\begin{equation}
\label{p2}
\begin{aligned}
P: \quad \textbf{H}^{\ \dot{\alpha}}_{1\ \dot{A}}&\longleftrightarrow\textbf{H}_{1A}^{\dagger{}\ \alpha}, \quad \textbf{H}_{2 \alpha}^{\ \ A}\longleftrightarrow\textbf{H}_{2\ \dot{\alpha}}^{\dagger{}\dot{A}},\\
\widetilde{\textbf{H}}_{1 \alpha}^{\ \ A}&\longleftrightarrow\widetilde{\textbf{H}}_{1\ \dot{\alpha}}^{\dagger{}\dot{A}}, \quad \widetilde{\textbf{H}}^{\ \dot{\alpha}}_{2\ \dot{A}}\longleftrightarrow\widetilde{\textbf{H}}_{2A}^{\dagger{}\ \alpha},\\
\bar{\textbf{H}}_{1 \alpha}^{\ \ A}&\longleftrightarrow \bar{\textbf{H}}_{1\ \dot{\alpha}}^{\dagger{}\dot{A}}, \quad \bar{\textbf{H}}^{\ \dot{\alpha}}_{2\ \dot{A}}\longleftrightarrow\bar{\textbf{H}}_{2A}^{\dagger{}\ \alpha}.
\end{aligned}
\end{equation}
In addition to these previous results (Formulas (\ref{pt1}) and (\ref{p2})), let us add two more requirements. The first requirement is that the gauge fields $W^{\mu}$ and $\mathcal{W}^{\mu}$ transform under parity transformation into each other as
 \begin{equation}
\begin{aligned}
P: \quad  W^{\mu}&\longleftrightarrow\mathcal -i\mathcal{W}^{\mu};
\end{aligned}
\end{equation}
the second requirement is, for the extended $U(1)_Y$ symmetry sector of the theory, the parity transformation switches the two different extended $U(1)_Y$ representations (see Sec.~\ref{112727} for detail) into each other:
\begin{equation}
\label{u1}
P: \quad U(2)_{T^0,\ \theta} \longleftrightarrow U(2)_{Y, \ T^0,\ T^3, \ \theta}.
\end{equation}
These two requirements are not invented out of thin air. The essence of these requirements are that, whatever before or after the parity transformation, only the $SU(2)_L$ doublets participates in the weak interaction; meanwhile, $SU(2)_L$ doublets always transform under the $U(2)_{T^0,\ \theta}$ group (with $e^{2iY\theta T^0}$ as its group element) and the $GW(2)_R$ doublets transform under the $U(2)_{Y, \ T^0,\ T^3, \ \theta}$ group (with $e^{2Yi\theta T^0-2i\theta T^3}$ as its group element). 
Then, implementing the  parity transformation given through (\ref{pt1})$\sim$(\ref{u1}) on the Lagrangian (\ref{lagrangian}), one obtains an astonishing result: even though only left spinors participate in the $SU(2)_L$ weak interaction (it is in this sense that parity symmetry is violated in the SM), Lagrangian (\ref{lagrangian}) is invariant under parity transformation, i.e., there is no parity violation. This feature of the theory, constructed by extending the SM to accommodate extended Weyl symmetry, provides new insights on the understanding  of the parity violation in the SM  \cite{Lee:1956qn}.
\end{itemize}

\section{A second scalar field}\label{sectionc}

The symmetry group of the SM now has been generalized as follows:
\begin{gather}
\label{groupextending}
\bigg\{SU(3)_c\odot GW(3)_c\bigg\}\times SU(2)_L\nonumber \\
\times
\left\{ \begin{aligned}
 &U(2)_{T^0, \theta} \qquad \ \ \ \ \  \text{for } \mathbf{L} \text{ and } \mathbf{Q}_L \\
GW(2)_R &\bigodot U(2)_{T^0, T^3, \theta}\ \   \text{for } \mathbf{R} \text{ and } \mathbf{Q}_R
\end{aligned}\right\} \times W(1).
\end{gather}
One can see from Sec.~\ref{jia} and Sec.~\ref{jiajia} that such generalization is not so straightforward since several important issues should be treated carefully in order to accommodate $GW(2)_R$ and $U(1)_Y$ symmetry into the theory consistently. Especially,  the Higgs doublet in the SM should be generalized from the original single doublet to two $2\times2$ Higgs matrices  $\textbf{H}_1$ and $\textbf{H}_2$ (see Sec.~\ref{jiajia} for more detail).
Each of them has $4$ DoFs, thus there are $8$ real DoFs in total. Taking advantage of the  $SU(2)_L\times GW(2)_R$ symmetry,  6 DoFs in $\textbf{H}_1$ and $\textbf{H}_2$ can be fixed completely and only two real DoFs remain. Compared with the fact that in the SM only one real scalar field (the Higgs field) in the Higgs doublet survives after the breaking of the $SU(2)_L$ symmetry, one can see that \textit{a second scalar field}  now has emerged by enlarging the SM to accommodate extended Weyl symmetry appropriately. 

The question is in what form these two scalar fields show themselves? One obvious clue from the SM to answer this question is that, after the SSB of the vacuum, there \textit{must} be a residual $U(1)_{EM}$ symmetry for electromagnetism. Then it can be checked that, if the form of the Higgs matrices $\textbf{H}_1$ and $\textbf{H}_2$ are
\begin{equation}
\label{1}
\textbf{H}_1=\begin{pmatrix} 0&0\\ 0&\phi+\varphi \end{pmatrix},\  \ \textbf{H}_2=\begin{pmatrix}0&0\\ 0&\phi-\varphi \end{pmatrix}
\end{equation}
with $\phi$ and $\varphi$ two real scalar fields, they will be unchanged under a residual symmetry transformation:\footnote{The residual $SU(2)_L$ transformation is denoted as $\begin{pmatrix} e^{-i\theta}&0\\ 0&e^{i\theta} \end{pmatrix}$ and operates on the Higgs matrices from the left to the right, see Formula (\ref{higgs}); the matrix $\begin{pmatrix} e^{i\theta}&0\\ 0&e^{-i\theta} \end{pmatrix}$ operating on the Higgs matrices from the right to the left comes from the combination of $U(2)_{T^0,\ \theta}$ and $U(2)_{Y, \ T^0,\ T^3, \ \theta}$ transformations on the Higgs matrices, see Formula (\ref{7272}).}
\begin{equation}
\label{higgsu12}
\begin{aligned}
\textbf{H}_1\rightarrow &\begin{pmatrix} e^{-i\theta}&0\\ 0&e^{i\theta} \end{pmatrix}\begin{pmatrix} 0&0\\ 0&\phi+\varphi \end{pmatrix}\begin{pmatrix} e^{i\theta}&0\\ 0&e^{-i\theta} \end{pmatrix}=\textbf{H}_1;\\
\textbf{H}_2\rightarrow &\begin{pmatrix} e^{-i\theta}&0\\ 0&e^{i\theta} \end{pmatrix}\begin{pmatrix} 0&0\\ 0&\phi-\varphi \end{pmatrix}\begin{pmatrix} e^{i\theta}&0\\ 0&e^{-i\theta} \end{pmatrix}=\textbf{H}_2.
\end{aligned}
\end{equation}
Furthermore, the $SU(2)_L$ doublets, $\textbf{Q}_L$ and $\textbf{L}$, transform under the residual $SU(2)_L\times U(2)_{T^0,\ \theta}$ transformation as
\begin{equation}
\label{spinoru1}
\begin{aligned}
\textbf{Q}_L\rightarrow &\begin{pmatrix} e^{-i\theta}&0\\ 0&e^{i\theta} \end{pmatrix} \begin{pmatrix} e^{-i\theta /3}&0\\ 0&e^{-i\theta /3} \end{pmatrix}\textbf{Q}_L\\
&\qquad \qquad =\begin{pmatrix} e^{-4i\theta/3}&0\\ 0&e^{2i\theta/3} \end{pmatrix}\textbf{Q}_L,\\
\textbf{L}\rightarrow &\begin{pmatrix} e^{-i\theta}&0\\ 0&e^{i\theta} \end{pmatrix}  \begin{pmatrix} e^{i\theta }&0\\ 0&e^{i\theta} \end{pmatrix}\textbf{L}=\begin{pmatrix} 1&0\\ 0&e^{2i\theta} \end{pmatrix}\textbf{L}.
\end{aligned}
\end{equation}
Formulas (\ref{4242}), (\ref{higgsu12}), and (\ref{spinoru1}) indicate that, while the two scalar fields and the neutrino (the upper components of the $SU(2)_L$ doublet $\textbf{L}$ and the $GW(2)_R$ doublet $\textbf{R}$) are electroneutral, the electron, the up quark and the down quark have electric charge as $-e$, $2e /3$ and $-e /3$ respectively \footnote{Here we follow the conventions in \cite{Cottingham:2007zz}.}, which are in good accordance with results in the SM. This also indicates that the two scalar fields should indeed show themselves in the way given by (\ref{1}). Then one of the two scalar fields can undertake the Weyl symmetry to protect the existence of the other scalar field (the Higgs field), just as what we expected from the beginning.  

The Higgs potential now takes the form as 
\begin{equation}
\label{123}
\begin{aligned}
V(\textbf{H}_1,\ \textbf{H}_2)&=\frac{1}{8}\mathrm{Tr}\Big{(}\textbf{H}_1^\dagger{}\textbf{H}_2\textbf{H}_1^\dagger{}\textbf{H}_2+\textbf{H}_2^\dagger{}\textbf{H}_1\textbf{H}_2^\dagger{}\textbf{H}_1 \Big{)}\\
&=\frac{1}{4}\big{(}\phi^2-\varphi^2\big{)}^2.
\end{aligned}
\end{equation}
Thus the vacuum locates at $\phi=\varphi$ where the potential has its minimum. A disturbing issue is that, after the SSB of the $SU(2)_L\times GW(2)_R\times U(1)_Y$ symmetry of the vacuum,  there are no masses for leptons and quarks. 
To see this, noticing that if the two Higgs matrices $\textbf{H}_1$ and $\textbf{H}_2$ are in the form given by Eq.(\ref{n9}),  the four induced Higgs matrices listed in Eq.(\ref{htilde}) and (\ref{hbar}) take the following form:
\begin{equation}
\label{hh}
\begin{aligned}
\widetilde{\textbf{H}}_1&=\begin{pmatrix} \phi+\varphi &0\\ 0&0\end{pmatrix},\  \ \ \bar{\textbf{H}}_1=\begin{pmatrix} \phi+\varphi&0\\ 0&\phi+\varphi \end{pmatrix},\\
\widetilde{\textbf{H}}_2&=\begin{pmatrix} \phi-\varphi &0\\ 0&0\end{pmatrix},\  \ \ \bar{\textbf{H}}_2=\begin{pmatrix} \phi-\varphi&0\\ 0&\phi-\varphi \end{pmatrix}.
\end{aligned}
\end{equation}
Then expanding $\phi$ around $\varphi$ (at where the vacuum locates) as $\phi=h+\varphi$ with $h$ the Higgs field, the Yukawa part in Lagrangian (\ref{lagrangian}) reads
\begin{equation}
\label{112233}
\begin{aligned}
\mathcal{L}_{\mathrm{Yukawa}}=&\frac{1}{2}\Big{(}\textbf{L}^{\dagger{}}(\textbf{H}_1+\bar{\textbf{H}}_2+\widetilde{\textbf{H}}_2)\textbf{R}+h.c.\Big{)} \\
&+\frac{1}{2}\Big{(}\textbf{Q}_L^{\dagger{}}(\textbf{H}_2+\bar{\textbf{H}}_1+\widetilde{\textbf{H}}_1)\textbf{Q}_R+h.c.\Big{)} \\
=&\nu_{L}^{\dagger{}}h\nu_{R}+\nu_{R}^{\dagger{}}h\nu_{L}\\
&+e_{L}^{\dagger{}}(h+\varphi)e_{R}+e_{R}^{\dagger{}}(h+\varphi)e_{L}\\
&+u_{L}^{\dagger{}}(h+2\varphi)u_R+u_{R}^{\dagger{}}(h+2\varphi)u_L\\
&+d_{L}^{\dagger{}}(h+\varphi)d_R+d_{R}^{\dagger{}}(h+\varphi)d_L.
\end{aligned}
\end{equation}
It can be seen that there are interactions between quarks and the two scalar fields but no mass for fermions in Eq.(\ref{112233}).

The way to solve this difficulty is to break the Weyl symmetry (i.e., $W(1)$ symmetry) to set one of the two scalar fields into a dimensional constant. Doing this can create connections between the theory and the real world. It should be emphasized that this is obligatory but not just a personal taste. The physical meaning of Weyl symmetry is that physics are independent of unit system. The meaning of breaking the Weyl symmetry is setting up a unit system to make measurements and statements about dimensional quantities in the real world having definite meaning. Otherwise it is always ambiguous to make statements about a dimensional quantity even in the simplest case. For example, the statement that "its mass is 1 kilogram" would be ambiguous and thus meaningless on a scientific level if the unit "kilogram" has not be well defined. It is in this sense that the Weyl symmetry has to be broken in order to relate the theory to the real world.
Then either $\phi$ or $\varphi$ can be transformed into a constant field by $W(1)$ symmetry. Without loss of generality,  one can treat $\varphi$ as so called \textit{a second scalar field} and set it to be a constant denoted as $v$. Then the Higgs matrices $\textbf{H}_1$ and $\textbf{H}_2$ are in the form of
\begin{equation}
\label{11}
\textbf{H}_1=\begin{pmatrix} 0&0\\ 0&\phi+v \end{pmatrix},\  \ \textbf{H}_2=\begin{pmatrix}0&0\\ 0&\phi-v \end{pmatrix}.
\end{equation}
Furthermore, the Higgs potential now is
\begin{align}
\label{1234}
V(\textbf{H}_1,\ \textbf{H}_2)=\dfrac{1}{4}\big{(}\phi^2-v^2\big{)}^2,
\end{align}
which is the same Higgs potential as that in the SM after the SSB of the vacuum. 

The afore mentioned disturbing issue that there are no masses for leptons and quarks after the SSB of the vacuum (see Eq.(\ref{112233})) is solved. To see this, using Eq.(\ref{11}) and expanding $\phi$ around $v$ (at where the vacuum now locates) as $\phi=h+v$ with $h$ the Higgs field, the Yukawa part of the Lagrangian (\ref{lagrangian}) now is
\begin{equation}
\label{lagragianyukawa}
\begin{aligned}
\mathcal{L}_{\mathrm{Yukawa}}=&\nu_{L}^{\dagger{}}h\nu_{R}+\nu_{R}^{\dagger{}}h\nu_{L}\\
&+e_{L}^{\dagger{}}(h+v)e_{R}+e_{R}^{\dagger{}}(h+v)e_{L}\\
&+u_{L}^{\dagger{}}(h+2v)u_R+u_{R}^{\dagger{}}(h+2v)u_L\\
&+d_{L}^{\dagger{}}(h+v)d_R+d_{R}^{\dagger{}}(h+v)d_L.
\end{aligned}
\end{equation}
The electron $e$, up quark $u$ and down quark $d$ all obtain masses. Those non-Abelian gauge particles also obtain masses through their couplings to the Higgs matrices in the kinetic terms. For the not yet observed right-handed neutrino $\nu_R$, term $(\nu_{L}^{\dagger{}}h\nu_{R}+\nu_{R}^{\dagger{}}h\nu_{L})$ indicates that the right-handed neutrino $\nu_R$ and the left-handed neutrino $\nu_L$ can convert into each other through the interaction with the Higgs field, which can be tested by future precise measurements in the Higgs sector. Therefore, it can be seen that , throughout the entire process, the SSB of the $SU(2)_L\times GW(2)_R\bigodot U(2)_{Y, \ T^0,\ T^3, \ \theta}\times W(1)$ invariant vacuum plays a pivotal role in making the emergence of \textit{a second scalar field}. Then accompanying  by breaking the $W(1)$ symmetry, this scalar field is transformed into a  dimensional constant which further leads to the mass generation of various fundamental particles. 

It is noteworthy that the SSB of the vacuum exerts an additional subtle yet critical influence: the SSB of the vacuum splits the $GW(2)_R\bigodot U(2)_{Y, \ T^0,\ T^3, \ \theta}$ group into two subgroups, spontaneously breaking one symmetry, with the residual symmetry corresponding to the electromagnetic gauge symmetry, thus allowing us to explicitly determine the specific form of the covariant derivative. Reminding that the group element of $GW(2)_R\bigodot U(2)_{Y, \ T^0,\ T^3, \ \theta}$ is $e^{(\eta_i(\theta, \ \omega)+i\tau_i(\theta, \ \omega))T^i}e^{2i\theta T^0}$ (see Formula (\ref{6161})). It seems that the corresponding covariant derivative of $GW(2)_R\bigodot U(2)_{Y, \ T^0,\ T^3, \ \theta}$ symmetry should take the form directly as
 \begin{equation}
 \mathscr{D}^{\mu}=\partial^{\mu}-(E^{\mu}_i+iF^{\mu}_i)T^i-iB^{\mu}T^0
 \end{equation}
 with $E^{\mu}_i$ and $F^{\mu}_i$ the six gauge fields corresponding to $\eta_i$ and $\tau_i$. However, it is in fact not so straightforward
since $\eta_i$ and $\tau_i$ are constructed by $\omega_i$ and $\theta$ carrying only only four independent DoFs in them, so the number of the gauge fields should be four but not six.
An effective approach to addressing this issue requires a comprehensive consideration of the implications arising from the SSB of the vacuum. The SSB in fact requires to factorize the group element $e^{(\eta_i(\theta, \ \omega)+i\tau_i(\theta, \ \omega))T^i}e^{2i\theta T^0}$ into the product of two different group elements as follows:
\begin{equation}
e^{(\eta_i(\theta, \ \omega)+i\tau_i(\theta, \ \omega))T^i}e^{2i\theta T^0}\longrightarrow e^{-\omega_i T^i}e^{2i\theta (T^0-T^3)}.
\end{equation}
Then, the $GW(2)_R$ part (related to $e^{-\omega_iT^i}$) will be broken, leaving the $U(2)_{Y, \ T^0,\ T^3, \ \theta}$ part (with $e^{2i\theta (T^0-T^3)}$ as its group element) as the residual $U(1)_{EM}$ group for the right-handed spinors. Therefore, from the perspective of the SSB of the vacuum, even though the $GW(2)_R\bigodot U(2)_{Y, \ T^0,\ T^3, \ \theta}$ group is not simply a direct product of $GW(2)_R$ and $U(2)_{Y, \ T^0,\ T^3, \ \theta}$, the SSB  will splits the original group into these two subgroups that have distinct  physical manifestations, indicating that the covariant derivative related to $GW(2)_R\bigodot U(2)_{Y, \ T^0,\ T^3, \ \theta}$ group can be expressed as (take the covariant derivative operated on $\textbf{R}$ for example)
\begin{equation}
\mathscr{D}^{\mu}\textbf{R}=(\partial^{\mu}+\mathcal{W}^{\mu}_iT^i-iB^{\mu}T^0+iB^{\mu}T^3)\textbf{R}.
\end{equation}
Other covariant derivatives related to $GW(2)_R\bigodot U(2)_{Y, \ T^0,\ T^3, \ \theta}$ symmetry can be given in the same way, see Formulas (\ref{d3}) $\sim$ (\ref{d7}).

Taking a close look at the whole process, there may be two unexpected bonuses which should be noticed.
\begin{itemize}

\item \textit{The sign problem in the Higgs mechanism}: While the Higgs potential in the present construction \big{(}see Eq.(\ref{lagrangian}), (\ref{123}) and (\ref{1234})\big{)} is constrained completely by $SU(2)_L\times GW(2)_R\times U(1)_Y\times W(1)$ symmetry and its broken, the process of obtaining the same Higgs potential in the SM differs significantly. In the SM, $SU(2)_L\times U(1)_Y$ symmetry can only constrain the Higgs potential to have the form as
\begin{equation}
V(\textbf{H})=\dfrac{1}{4}\big{(}\phi^2 \pm v^2\big{)}^2,
\end{equation}
with both plus and minus signs in the front of $v^2$ can meet the requirements of $SU(2)_L\times U(1)_Y$ symmetry. Then, \textit{in order to} spontaneously break the symmetry of the vacuum, the sign before $v^2$ is set \textit{by hand} to be negative. This is why it is believed that understanding the Higgs mechanism completely needs a more fundamental answer to the question that why the parameter before $v^2$ is negative \cite{Peskin:2015kka}.
\ Since in the present construction the minus sign in the front of $v^2$ is a natural result of the Weyl and extended Weyl symmetry extension of the SM and symmetry broken, not set by hand, the present work gives a natural and fundamental explanation to the sign problem in the Higgs mechanism.

\item \textit{The naturalness of the small neutrino mass}: Astonishingly, even though the neutrino couples to the Higgs field and possesses both left- and right-handed components (in the SM this would imply a non-zero mass), it remains strictly massless in the present framework. Unlike the minimal SM---where the absence of the right-handed neutrino $\nu_R$ and the resulting zero neutrino mass are imposed \textit{by hand}---the inclusion of $\nu_R$ and the vanishing neutrino mass here are required and induced by the extended Weyl symmetry and its subsequent breaking. Admittedly, neutrino oscillation experiments~\cite{Super-Kamiokande:1998kpq,SNO:2001kpb} demonstrate that neutrinos do have non-zero masses, so the strict masslessness may appear incompatible with observation. However, the observed masses are not only non-zero but also extraordinarily small compared to those of other SM fermions. According to 't Hooft's naturalness criterion~\cite{thooft}, such a small mass is natural only if setting it to zero enhances the symmetry of the theory. This is naturally realized in the present framework at tree level: the extended Weyl symmetry enforces a massless neutrino. In other words, the smallness of the neutrino mass is protected by the extended Weyl symmetry. Therefore, the extended Weyl extension of the SM offers a natural explanation of the puzzling smallness of the neutrino mass.
\end{itemize}

\section{Summary and discussion}
This work establishes an equivalent yet more compact reformulation of the SM, in which the Higgs doublet is recast as a $2\times2$ matrix field and right-handed fermionic singlets are assembled into novel doublets. This structure reveals two intrinsic avenues for extension: first, the Higgs matrix naturally accommodates additional scalar DoFs beyond its original column-vector form; second, although right-handed fermions are inert under weak interactions, their organization into doublets suggests the possibility of new symmetries that extends the scope of local gauge symmetries beyond the conventional Yang--Mills framework.

Guided by these insights, we revisit Hermann Weyl's original notion of gauge symmetry and construct its generalization: the \textbf{extended Weyl symmetry}. Denoted $GW(N)$, this symmetry represents a fundamental extension of Weyl's 1929 idea of local scale invariance to internal multiplet spaces. Unlike Yang--Mills theories, which arise from local phase rotations, extended Weyl symmetry emerges from local rescalings intertwined with internal algebraic structure. Its construction constitutes a novel gauge principle in its own right---one that would merit attention even in the absence of the SM. Remarkably, the reformulated SM not only accommodates but actively realizes this symmetry: the right-handed doublets become natural carriers of a $GW(2)_R$ gauge group, while the enlarged Higgs sector provides the necessary scalar degrees of freedom to protect the physical Higgs boson.

Building on this foundation, we develop a classical extension of the SM based on extended Weyl symmetry. The extension of color symmetry from $SU(3)_c$ to $SU(3)_c \odot GW(3)_c$ proceeds straightforwardly (see Sec.~\ref{su3}), whereas the electroweak sector presents a nontrivial challenge. To respect experimental constraints---specifically, the absence of right-handed couplings to $SU(2)_L$ gauge bosons---we require that $GW(2)_R$ act exclusively on right-handed spinors. This leads to the symmetry embedding
\begin{equation}
SU(2)_L \longrightarrow SU(2)_L \times GW(2)_R,
\end{equation}
under which the right-handed up and down quarks form a $GW(2)_R$ doublet. Consistency of this construction \textit{demands} the existence of a right-handed neutrino $\nu_R$, which pairs with the right-handed electron $e_R$ to complete a second $GW(2)_R$ doublet. The Higgs sector is correspondingly enlarged from a single doublet to two $2\times2$ matrix fields, $\mathbf{H}_1$ and $\mathbf{H}_2$, which serve as bridges between left- and right-handed sectors. Moreover, because the components of each $GW(2)_R$ doublet carry different hypercharges under $U(1)_Y$, the embedding necessitates a refinement of the hypercharge assignment, endowing $U(1)_Y$ with richer algebraic meaning. With these elements consistently arranged, the full classical action of the Weyl-extended SM is obtained (see Secs.~\ref{jia} and~\ref{jiajia}).

This extension yields several concrete benefits. First, it predicts new gauge bosons associated with $GW(3)_c$ and $GW(2)_R$. The $GW(3)_c$ gauge bosons decouple entirely from all SM fields---including the Higgs sector---and thus constitute a secluded dark matter candidate accessible only through gravitational or cosmological signatures. In contrast, the $GW(2)_R$ gauge bosons couple exclusively to the extended Higgs matrices $\mathbf{H}_1$ and $\mathbf{H}_2$, rendering them invisible to conventional weak-interaction probes but potentially detectable via precision measurements of Higgs couplings (see Secs.~\ref{su3} and~\ref{3.4}). Both sectors therefore offer distinct dark matter scenarios: one completely hidden, the other indirectly testable.

Second, the framework successfully addresses the problem we set out to solve at the outset: how to consistently introduce a second scalar field to neutralize the Weyl symmetry and thereby protect the physical Higgs boson from being gauged away. In this construction, two real scalar degrees of freedom arise naturally from the $2\times2$ Higgs matrix structure. After spontaneous symmetry breaking of the vacuum and the breaking of the $W(1)$ symmetry, one of these scalars acquires a vacuum expectation value and becomes non-dynamical---effectively fixing the Weyl scale as a constant background---while the other remains dynamical and manifests as the observed Higgs field, generating masses for quarks and charged leptons (see Sec.~\ref{sectionc}).

Astonishingly, beyond these anticipated results, the extension also yields several unexpected theoretical insights.
First, it reconciles the apparent parity violation of the SM with an underlying parity-symmetric structure. Although only left-handed fermions participate in $SU(2)_L$ interactions---leading to manifest chiral asymmetry---the full theory possesses a hidden parity symmetry: the parity transformation interchanges the $SU(2)_L$ and $GW(2)_R$ sectors, mapping left-handed fields to right-handed ones and vice versa. The observed parity violation thus stems not from a fundamental asymmetry of the laws of physics, but from the phenomenological fact that the $GW(2)_R$ gauge bosons couple exclusively to the Higgs sector and remain invisible to conventional weak probes. Only through high-precision measurements of Higgs couplings can this hidden parity symmetry be experimentally revealed.

Second, the framework provides a fundamental explanation for the so-called sign problem in the Higgs mechanism. In the Standard Model, the $SU(2)_L \times U(1)_Y$ symmetry alone constrains the Higgs potential only up to the form
\begin{equation}
V(\mathbf{H}) \propto \bigl(\phi^2 \pm v^2\bigr)^2,
\end{equation}
and the minus sign must be chosen \textit{by hand} to allow spontaneous symmetry breaking. However, when the SM is extended to incorporate local Weyl and extended Weyl symmetries---and these symmetries are subsequently broken in the appropriate pattern---the scalar potential is uniquely fixed to the form
\begin{equation}
V(\mathbf{H}_1,\mathbf{H}_2) \propto \bigl(\phi^2 - v^2\bigr)^2,
\end{equation}
with the negative sign arising necessarily from the symmetry structure (see Sec.~\ref{sectionc}). Thus, the ambiguity disappears once the role of Weyl and extended Weyl symmetries is recognized.

Third, the model provides a compelling rationale for the existence of the right-handed neutrino: it is not an optional addition, but a mandatory component required by the $GW(2)_R$ doublet structure. Moreover, the neutrino couples to the Higgs matrices, offering a potential avenue for its indirect detection through future precision Higgs measurements. In typical Yukawa scenarios, such a coupling would imply a non-zero mass after electroweak symmetry breaking. Remarkably, in the present framework, the neutrino couples to the Higgs matrices yet remains strictly massless (see Eq.~\eqref{lagragianyukawa}). This offers a natural explanation for the puzzling smallness of neutrino masses: the vanishing mass is protected by extended Weyl symmetry, and setting it to zero enhances the symmetry of the theory---precisely realizing the extended Weyl symmetry.

Several directions warrant further investigation. The quantum consistency of extended Weyl symmetry---including potential anomalies, renormalizability, and dynamical scale generation---remains an open challenge. Additionally, the detailed phenomenology of the $GW(3)_c$ and $GW(2)_R$ gauge bosons, as well as mechanisms for generating realistic neutrino masses within this framework, deserve dedicated study. Nevertheless, this work lays essential classical groundwork: while the reformulation of the SM serves as the natural starting point, the central theoretical contribution lies in the proposal of extended Weyl symmetry---a new class of local gauge transformations that enriches the very architecture of modern field theory. It demonstrates that the SM's latent algebraic structure not only permits but actively invites a consistent extension along this novel symmetry direction, transforming seemingly ad hoc model-building into a theoretically inevitable consequence and revealing that the Standard Model, when viewed through its latent algebraic geometry, already contains the seeds of its own consistent enlargement.


\begin{thebibliography}{99}

\bibitem{Glashow:1961tr}
S.~L.~Glashow,
``Partial Symmetries of Weak Interactions,''
Nucl. Phys. \textbf{22} (1961), 579-588
doi:10.1016/0029-5582(61)90469-2

\bibitem{Weinberg:1967tq}
S.~Weinberg,
``A Model of Leptons,''
Phys. Rev. Lett. \textbf{19} (1967), 1264-1266
doi:10.1103/PhysRevLett.19.1264

\bibitem{Salam:1968rm}
A.~Salam,
``Weak and Electromagnetic Interactions,''
Conf. Proc. C \textbf{680519} (1968), 367-377
doi:10.1142/9789812795915{\_}0034


\bibitem{Weyl:1929fm}
H.~Weyl,
``Electron and Gravitation. 1. (In German),''
Z. Phys. \textbf{56} (1929), 330-352
doi:10.1007/BF01339504

\bibitem{Yang:1954ek}
C.~N.~Yang and R.~L.~Mills,
``Conservation of Isotopic Spin and Isotopic Gauge Invariance,''
Phys. Rev. \textbf{96} (1954), 191-195
doi:10.1103/PhysRev.96.191

\bibitem{Englert:1964et}
F.~Englert and R.~Brout,
``Broken Symmetry and the Mass of Gauge Vector Mesons,''
Phys. Rev. Lett. \textbf{13} (1964), 321--323
doi:10.1103/PhysRevLett.13.321

\bibitem{Higgs:1964pj}
P.~W.~Higgs,
``Broken Symmetries and the Masses of Gauge Bosons,''
Phys. Rev. Lett. \textbf{13} (1964), 508--509
doi:10.1103/PhysRevLett.13.508

\bibitem{Guralnik:1964eu}
G.~S.~Guralnik, C.~R.~Hagen and T.~W.~B.~Kibble,
``Global Conservation Laws and Massless Particles,''
Phys. Rev. Lett. \textbf{13} (1964), 585--587
doi:10.1103/PhysRevLett.13.585

\bibitem{Discovery}
ATLAS Collaboration, Phys.Lett. B716 (2012) 1-29.\\	
CMS Collaboration, Phys. Lett. B 716 (2012) 30

\bibitem{Cheng:1988zx}
H.~Cheng,
``The Possible Existence of Weyl's Vector Meson,''
Phys. Rev. Lett. \textbf{61} (1988), 2182
doi:10.1103/PhysRevLett.61.2182

\bibitem{Cheng:1988jy}
H.~Cheng and W.~F.~Kao,
``CONSEQUENCES OF SCALE INVARIANCE,''
Print-88-0907 (MIT).

\bibitem{Ghilencea:2021lpa}
D.~M.~Ghilencea,
``Standard Model in Weyl conformal geometry,''
Eur. Phys. J. C \textbf{82} (2022) no.1, 23
doi:10.1140/epjc/s10052-021-09887-y
[arXiv:2104.15118 [hep-ph]].

\bibitem{Ghilencea:2023sti}
D.~M.~Ghilencea,
``Weyl conformal geometry vs Weyl anomaly,''
JHEP \textbf{10} (2023), 113
doi:10.1007/JHEP10(2023)113
[arXiv:2309.11372 [hep-th]].

\bibitem{Meissner:2006zh}
K.~A.~Meissner and H.~Nicolai,
``Conformal Symmetry and the Standard Model,''
Phys. Lett. B \textbf{648} (2007), 312-317
doi:10.1016/j.physletb.2007.03.023
[arXiv:hep-th/0612165 [hep-th]].

\bibitem{Quiros:2014wda}
I.~Quiros,
``Scale invariance: fake appearances,''
[arXiv:1405.6668 [gr-qc]].

\bibitem{Ghilencea:2023wwf}
D.~M.~Ghilencea and C.~T.~Hill,
``Standard Model in conformal geometry: Local vs gauged scale invariance,''
Annals Phys. \textbf{460} (2024), 169562
doi:10.1016/j.aop.2023.169562
[arXiv:2303.02515 [hep-th]].

\bibitem{Singh:2011av}
N.~K.~Singh, P.~Jain, S.~Mitra and S.~Panda,
``Quantum Treatment of the Weyl Vector Meson,''
Phys. Rev. D \textbf{84} (2011), 105037
doi:10.1103/PhysRevD.84.105037
[arXiv:1106.1956 [hep-ph]].

\bibitem{Kashyap:2012kg}
G.~Kashyap,
``Weyl meson and its implications in collider physics and cosmology,''
Phys. Rev. D \textbf{87} (2013) no.1, 016018
doi:10.1103/PhysRevD.87.016018
[arXiv:1207.6195 [hep-ph]].

\bibitem{Tang:2019uex}
Y.~Tang and Y.~L.~Wu,
``Weyl Symmetry Inspired Inflation and Dark Matter,''
Phys. Lett. B \textbf{803} (2020), 135320
doi:10.1016/j.physletb.2020.135320
[arXiv:1904.04493 [hep-ph]].

\bibitem{Hayashi:1976uz}
K.~Hayashi, M.~Kasuya and T.~Shirafuji,
``Elementary Particles and Weyl's Gauge Field,''
Prog. Theor. Phys. \textbf{57} (1977), 431
[erratum: Prog. Theor. Phys. \textbf{59} (1978), 681]
doi:10.1143/PTP.57.431


\bibitem{Muller-Kirsten:1986ysr}
H.~J.~W.~Muller-Kirsten and A.~Wiedemann,
``SUPERSYMMETRY: AN INTRODUCTION WITH CONCEPTUAL AND CALCULATIONAL DETAILS,''
PRINT-86-0955.

\bibitem{Robinson:2011lia}
M.~Robinson,
``Symmetry and the standard model: Mathematics and particle physics,''
Springer, 2011,
ISBN 978-1-4419-8266-7
doi:10.1007/978-1-4419-8267-4

\bibitem{Schwichtenberg:2018dri}
J.~Schwichtenberg,
``Physics from Symmetry,''
Springer International Publishing, 2018,
ISBN 978-3-319-66630-3, 978-3-319-66631-0
doi:10.1007/978-3-319-66631-0


\bibitem{Lee:1956qn}
T.~D.~Lee and C.~N.~Yang,
``Question of Parity Conservation in Weak Interactions,''
Phys. Rev. \textbf{104} (1956), 254-258
doi:10.1103/PhysRev.104.254

\bibitem{Cottingham:2007zz}
W.~N.~Cottingham and D.~A.~Greenwood,
``An Introduction to the Standard Model of Particle Physics,''
Cambridge University Press, 2023,
ISBN 978-1-009-40168-5, 978-1-009-40172-2, 978-1-009-40170-8, 978-0-511-27136-6, 978-0-521-85249-4
doi:10.1017/9781009401685

\bibitem{Peskin:2015kka}
M.~E.~Peskin,
``On the Trail of the Higgs Boson,''
Annalen Phys. \textbf{528} (2016) no.1-2, 20-34
doi:10.1002/andp.201500225
[arXiv:1506.08185 [hep-ph]].

\bibitem{Super-Kamiokande:1998kpq}
Y.~Fukuda \textit{et al.} [Super-Kamiokande],
``Evidence for oscillation of atmospheric neutrinos,''
Phys. Rev. Lett. \textbf{81} (1998), 1562-1567
doi:10.1103/PhysRevLett.81.1562
[arXiv:hep-ex/9807003 [hep-ex]].

\bibitem{SNO:2001kpb}
Q.~R.~Ahmad \textit{et al.} [SNO],
``Measurement of the rate of $\nu_e+d \to p+p+e^-$ interactions produced by $^8$B solar neutrinos at the Sudbury Neutrino Observatory,''
Phys. Rev. Lett. \textbf{87} (2001), 071301
doi:10.1103/PhysRevLett.87.071301
[arXiv:nucl-ex/0106015 [nucl-ex]].

\bibitem{thooft}
G.~'t Hooft, in Proc. of 1979 Carg\`ese Institute on {\it Recent Developments in Gauge Theories}, p.~135, Plenum Press, New York 1980.





\end{thebibliography}
\end{document}